%% file: cas-dc-sample.tex
\documentclass[a4paper,fleqn]{cas-dc}

\usepackage[numbers]{natbib}
\usepackage{enumitem}

\usepackage{tikz}
\usepackage{forest}       
\useforestlibrary{edges} 
\usepackage{xcolor}       
\usepackage{adjustbox}
\usepackage{placeins}

\def\tsc#1{\csdef{#1}{\textsc{\lowercase{#1}}\xspace}}
\tsc{WGM}
\tsc{QE}
\tsc{EP}
\tsc{PMS}
\tsc{BEC}
\tsc{DE}

\begin{document}
\let\WriteBookmarks\relax



\makeatletter
\renewcommand{\topfraction}{0.95}      
\renewcommand{\bottomfraction}{0.95}    
\renewcommand{\textfraction}{0.01}      
\renewcommand{\floatpagefraction}{0.75} 
\renewcommand{\dbltopfraction}{0.95}    
\renewcommand{\dblfloatpagefraction}{0.75} 

\setcounter{topnumber}{4}
\setcounter{bottomnumber}{4}
\setcounter{totalnumber}{8}
\setcounter{dbltopnumber}{4}            

\setlength{\textfloatsep}{10pt plus 2pt minus 4pt}
\setlength{\floatsep}{8pt plus 2pt minus 2pt}
\setlength{\intextsep}{10pt plus 2pt minus 4pt}
\setlength{\dbltextfloatsep}{10pt plus 2pt minus 4pt}
\setlength{\dblfloatsep}{8pt plus 2pt minus 2pt}

\setlength{\@fptop}{0pt}
\setlength{\@fpbot}{0pt plus 1fil}
\setlength{\@fpsep}{8pt plus 2fil}
\setlength{\@dblfptop}{0pt}
\setlength{\@dblfpbot}{0pt plus 1fil}
\setlength{\@dblfpsep}{8pt plus 2fil}
\makeatother

\shorttitle{A Survey on Generative Recommendation}
\shortauthors{Min Hou et~al.}

\title [mode = title]{A Survey on Generative Recommendation: Data, Model, and Tasks}              

\author[1]{Min Hou}
\ead{hmhoumin@gmail.com}

\credit{Conceptualization of this study, Methodology, Software}

\affiliation[1]{organization={Hefei University of Technology},
                city={Hefei},
                postcode={230009}, 
                state={Anhui},
                country={China}}

\author[1]{Le Wu}[style=chinese]
\cormark[1]
\ead{lewu.ustc@gmail.com}

\author[1]{Yuxin Liao}[%
   ]
\ead{yuxinliao314@gmail.com}

\credit{Data curation, Writing - Original draft preparation}

\affiliation[2]{organization={National University of Singapore},
                country={Singapore}}

\author[2]{Yonghui Yang}
\ead{yyh.hfut@gmail.com}

\author[1]{Zhen Zhang}
\ead{zhangz.199911@gmail.com}

\author[1]{Yu Wang}
\ead{wangyu20001162@gmail.com}

\author[1]{Changlong Zheng}
\ead{changlongzheng419@gmail.com}

\author[1]{Han Wu}
\ead{ustcwuhan@gmail.com}

\author[1]{Richang Hong}
\ead{hongrc.hfut@gmail.com}

\cortext[cor1]{Corresponding author}


\begin{abstract}
Recommender systems serve as a foundational infrastructure in modern information ecosystems, helping users navigate the expanding digital content space and discover items aligned with their preferences. At their core, recommender systems address a fundamental research problem: matching users with items. Over the past decades, the field has experienced successive technological paradigm shifts, from collaborative filtering and matrix factorization in the machine learning era to sophisticated neural architectures in the deep learning era. Recently, the emergence of generative models, especially large language models (LLMs) and diffusion models have sparked a new paradigm: generative recommendation, which reconceptualizes the recommendation problem as a generation task rather than a discriminative scoring procedure. This survey provides a comprehensive examination of this paradigm through a unified tripartite framework spanning data, model, and task dimensions. Rather than simply categorizing works, we systematically decompose approaches into operational stages—data augmentation and unification, model alignment and training, task formulation and execution. At the data level, generative models enable knowledge-infused augmentation and agent-based simulation while unifying heterogeneous signals. At the model level, we taxonomize LLM-based methods, large recommendation models, and diffusion approaches, analyzing their alignment mechanisms and innovations. At the task level, we illuminate new capabilities including conversational interaction, explainable reasoning, and personalized content generation.
We identify five key advantages: world knowledge integration, natural language understanding, reasoning capabilities, scaling laws, and creative generation. We critically examine challenges in benchmark design, model robustness, and deployment efficiency, while charting a roadmap toward intelligent recommendation assistants that fundamentally reshape human-information interaction.
\end{abstract}


\maketitle

\input{introduction}
\input{opportunities}

\input{data_lyx}
\input{model}

\input{task}

\input{challenges}

\input{conclusion}
\input{acknowledgment}

\bibliographystyle{cas-model2-names}

\bibliography{cas-refs}

\end{document}

%% file: introduction.tex
\section{Introduction}
Recommender systems~(RSs) aim to recommend the items (e.g., E-commerce products, micro-videos, musics, and news) by inferring user interest from historical interactions, user profiles, and context information. 
RSs serve as an effective solution to alleviate information overload, to facilitate users seeking desired information, and to increase the traffic and revenue of service providers. 
Currently, RSs are extensively deployed across diverse domains, including e-commerce, social media, education, online video, and music services, establishing them as one of the most pervasive user-centered AI applications in modern information systems.

Over the past decade, RSs have made significant advancements, largely driven by improvements in deep learning architectures and the growing availability of large-scale user behavior data.  Traditional RSs can be viewed as a discriminative matching paradigm, focusing on learning an effective matching function between users and items. In the 1990s, early research developed heuristics for content-based and collaborative filtering methods~\cite{resnick1994grouplens}. In the 2000s, Matrix Factorization~\cite{koren2009matrix} gained prominence, especially after the Netflix Prize, marking a key milestone in the development of RSs. By the mid-2010s, advancements in neural networks, such as CNNs, RNNs, GNNs, and Transformers, led to a transition toward deep learning-based recommendation methods. These approaches take advantage of deep learning’s powerful representation capabilities to handle complex data, enabling non-linear mapping of user-item interactions and enhancing user/item representation learning by encoding intricate semantics (e.g., text, images, social networks, and knowledge graphs)~\cite{accuracy-oriented-survey-TKDE-2022}. Despite these advancements, traditional RSs still face several challenges. They remain reliant on limited semantic knowledge, often manually processed, and struggle with sub-optimal performance for small-scale models. Additionally, they depend on fixed candidate sets, require task-specific architectures and training objectives, encounter difficulties in cold-start scenarios, and struggle to provide transparent, context-rich explanations.

\begin{figure}[]
    \centering \includegraphics[width=0.48\textwidth]{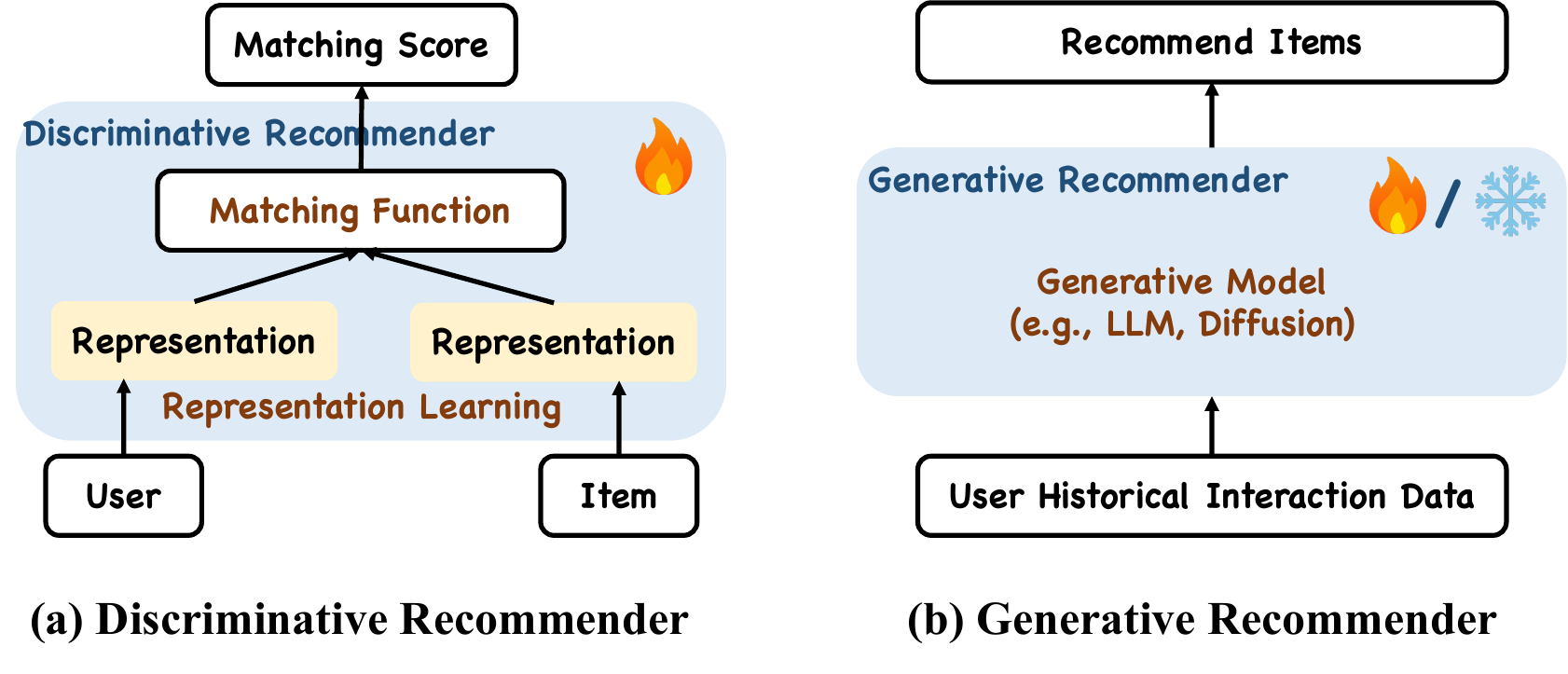}
    \caption{Discriminative Recommendation and Generative Recommendation}
    \label{fig:difference}
\end{figure}

Recently, generative models have made significant advances in artificial intelligence. In particular, Large Language Models (LLMs) and diffusion models have brought about revolutionary changes in AI-generated content (AIGC). Leveraging vast pre-training corpora and enhanced computational power, LLMs demonstrate emergent abilities, including in-context learning, complex reasoning, and so on. Meanwhile, diffusion models have led to remarkable breakthroughs in photo-realistic image generation.
In recent years, the emergence of generative models has brought about a paradigm shift in the research and development of RSs. Distinguished from previous discriminative matching paradigms, generative recommendation aims to directly generate the target document or item to satisfy users' information needs as shown in Fig. \ref{fig:difference}.
Embracing generative recommendation brings new benefits and opportunities for recommendation. In particular:
(1) LLMs inherently possess formidable capabilities, such as vast world knowledge, semantic understanding, interactive skills, and instruction following. These inherent abilities can be leveraged to augment sparse recommendation data, enrich item representations, simulate user behaviors, and unify heterogeneous information across domains and modalities, thereby addressing long-standing data sparsity and cold-start challenges.
(2) The tremendous success of LLMs stems from their generative learning paradigm and scaling laws. Applying generative learning to recommendation fundamentally revolutionizes the modeling approaches—from discriminative scoring to generative synthesis—enabling powerful architectures that can capture complex user-item interactions and demonstrate emergent capabilities through increased scale.
(3) LLMs-based generative AI applications, such as ChatGPT, are gradually becoming a new gateway for users to access Web content. Developing generative recommendation enables new task paradigms including conversational interaction, explainable reasoning, and personalized content generation, which can be seamlessly integrated into these generative AI applications.
To advance this rapidly evolving field, it is imperative to undertake a comprehensive survey of existing studies, building a technique map to systematically guide the research, practice, and service in generative model-empowered recommendation.

In this survey, we provide a comprehensive review of how generative models reshape recommendation. This survey organizes existing studies into three complementary perspectives: \textbf{data-level}, \textbf{model-level}, and \textbf{task-level}. This taxonomy reflects both the comprehensive influence of generative modeling across the entire recommendation pipeline and provides a systemic framework for understanding recent advances.

From a methodological perspective, the generative capability can be incorporated into recommendation processes at different stages. Data-level approaches focus on leveraging generative models to enhance or augment the data space—such as synthesizing user behaviors, generating item attributes, or mitigating data sparsity. Model-level approaches integrate generative mechanisms directly into the recommendation architecture. Task-level approaches further extend generative modeling to high-level objectives, including personalized content creation, explainable recommendation, and conversational recommendation tasks, etc.
This taxonomy not only delineates the scope of generative recommendation along the end-to-end system pipeline—ranging from data preparation to model design and task realization—but also echoes the historical evolution of the field.
Early research primarily employed generative modeling for data-level enhancement, addressing challenges such as data sparsity and bias through synthetic interactions or item generation.
Subsequent works advanced toward model-level innovation, where generative inference and probabilistic reasoning became integral parts of recommendation architectures.
More recently, task-level generation has emerged, extending generative capabilities to high-level applications such as personalized content creation, multi-modal recommendation, and explainable reasoning.
Together, these three perspectives form a coherent and progressive framework that reflects both the methodological depth and practical breadth of generative recommendation research—from its early data augmentation roots to its modern role as an intelligent, generation-driven paradigm.

\subsection{How Do We Collect the Papers?}
We collected over 200 papers related to generative models (LLMs and diffusion models) for recommender systems. Initially, we searched top-tier conferences and journals such as ICML, ICLR, NeurIPS, ACL, SIGIR, KDD, WWW, RecSys, TKDE, TOIS to identify related works in the past 3 years. We also include some highly cited work in ArXiv. To collect relevant literature, we first defined two groups of search keywords. The first group is related to recommendation systems, including terms such as ``recommendation", ``recommender system", and ``user modeling". The second group focuses on generative and foundation models, including “generative”, “agent”, “large language model”, “reasoning”, “foundation model”, and “retrieval-augmented generation (RAG)”. We then constructed combinations of these keywords to perform systematic searches.

\subsection{Contributions of This Survey}
Research on generative recommendation is accelerating. A growing body of surveys has recently emerged to review this emerging research direction. In 2024 and before, Wu et al. \cite{survey-llmrec-ustc} systematically reviewed the LLM-based recommendation systems. From the perspective of modeling paradigms, they categorized the studies into LLM Embeddings enhanced RS, LLM Tokens enhanced RS, and LLM as RS. Lin et al. \cite{survey-llmrec-huawei} introduced two orthogonal perspectives: where and how to adapt LLMs in recommender systems. They introduced the existing methods according to whether to tune LLM and whether to involve conventional recommendation models. Zhao et al. \cite{survey-llmrec-polyu} reviewed of LLM-empowered recommender systems from various
aspects, including pre-training, fine-tuning, and prompting paradigms. Deldjoo et al. \cite{survey-llmrec-ucsd} connects key advancements
in RS using Generative Models (Gen-RecSys), covering: interaction-driven generative models; the use of LLM and textual data for natural language recommendation; and the integration of multimodal models for generating and processing images/videos in RS. Liu et al. \cite{survey-llmrec-huawei-multimodal} explored the advancements in multimodal pretraining, adaptation, and generation techniques, as well as their applications to recommender systems. Li et al. \cite{survey-llmrec-hkbu} reviewed the recent progress of LLM-based generative recommendation and provided a general formulation for each generative recommendation task according to relevant research. Wang et al. \cite{survey-llmrec-jlu} reviewed existing LLM-based recommendation works and discussed the gap from academic research to industrial application.
\FloatBarrier
However, despite their comprehensive scope, these surveys primarily reflect the state of the field up until 2024, and many of them have overlooked the latest research that emerged in 2025 and beyond. Specifically, they fail to fully address the new research on agent-based recommender systems using LLMs, the rapidly growing exploration of LLM-based recommendation methods beyond supervised fine-tuning (SFT), and so on.
\begin{figure*}[!t]
    \centering \includegraphics[width=0.98\textwidth]{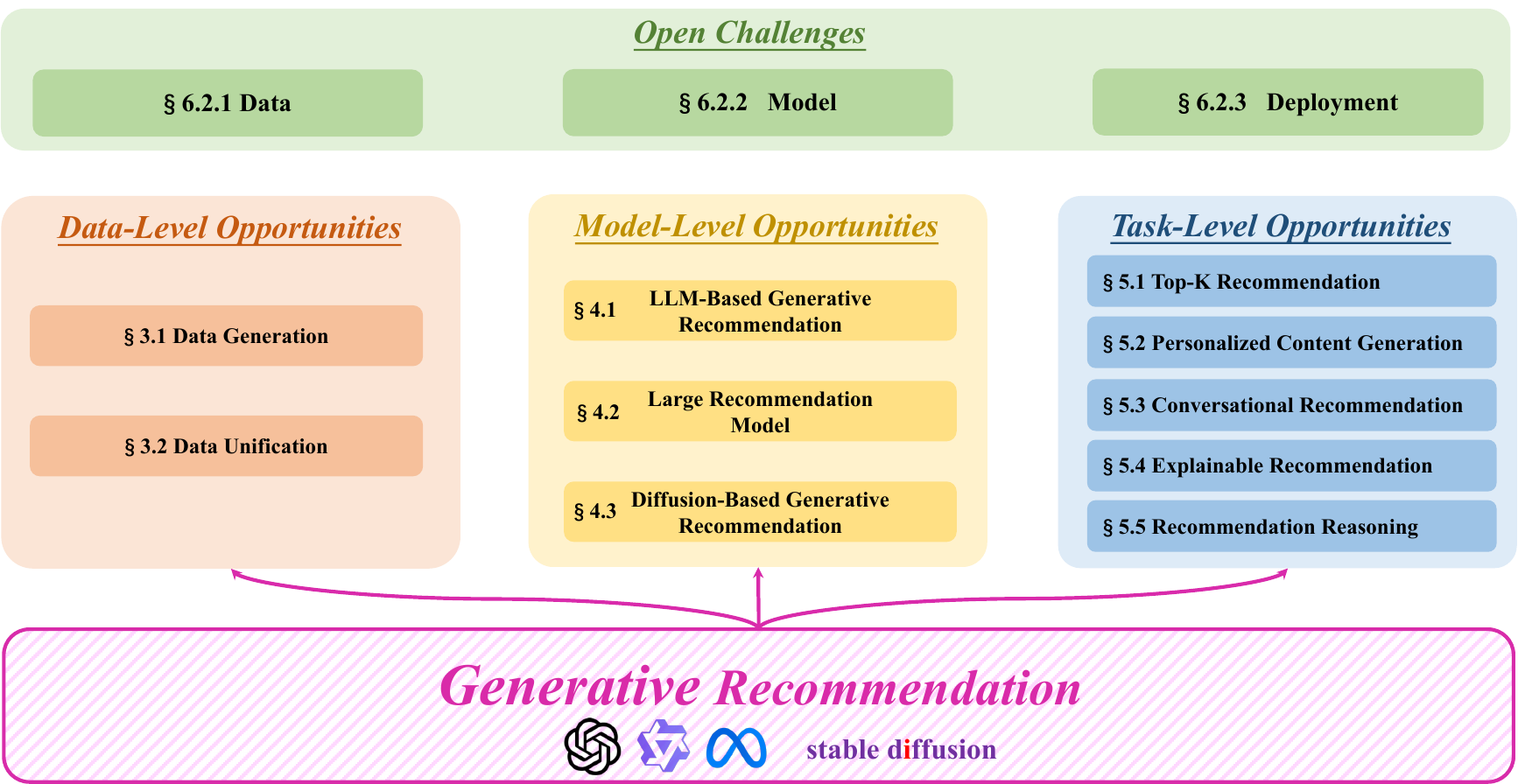}
    \caption{Overview of this survey.}
    \label{fig:overview}
\end{figure*}
In contrast, our survey provides three key contributions. First, we offer a broader coverage of generative paradigms, categorizing research into LLM-based generative recommendation, large recommendation models (LRMs), and diffusion-based generative recommendation, ensuring we incorporate the most recent advancements that may have been overlooked in earlier surveys. Second, we introduce a data–model–task framework, which allows us to analyze the contributions of generative models at different levels, from data preparation to model architecture to task-specific innovations. This framework provides a more comprehensive understanding of the evolving role of generative models in recommendation systems. Third, we dedicate a section to task-level innovations, examining how generative models unlock novel recommendation scenarios such as interactive recommendation, conversational recommendation, and personalized content generation—an area that has been underexplored in prior works. Finally, our survey concludes with a discussion on current challenges and future research directions, offering an up-to-date roadmap for advancing the field of generative recommendation systems, especially in light of recent developments that have yet to be fully explored in the literature.

\subsection{Scope and Organization of This Survey}
The landscape of recommendation modeling can be broadly examined across three fundamental dimensions: \textbf{data}, \textbf{model}, and \textbf{task}. This survey centers on the transformative impact of generative models on each of these dimensions, aiming to provide a systematic and comprehensive review of the current state of research. The overview of this survey is shown in Fig. \ref{fig:overview}.

The remainder of this paper is organized as follows.
\textbf{Section} \ref{sec2:Foundations of Generative Recommendation} introduces the preliminary and background of recommendation, contrasting traditional discriminative approaches with generative paradigms, and highlighting their key concepts, differences, and distinctive characteristics.
\textbf{Section} \ref{sec3:Data-Level Opportunities} surveys methods that incorporate generative models (particularly LLMs) at the data level, including feature generation, interaction synthesis, and data integration.
\textbf{Section} \ref{sec4:Model-Level Opportunities} examines approaches that leverage generative models as the core recommendation engine. We categorize mainstream research into three directions: LLM-based generative recommendation, large recommendation models, and diffusion-based generative recommendation.
\textbf{Section} \ref{sec5:Task and Application-Level Opportunities} discusses task-level innovations enabled by generative models in recommendation.
\textbf{Section} \ref{sec6:Open Challenges} outlines open challenges and future research opportunities in generative recommendation, proposing directions for advancing this emerging field.
Finally, \textbf{Section} \ref{sec7:Conclusion} concludes the survey by summarizing the contributions of generative models to RSs.

\input{fig_tree}


%% file: fig_tree.tex


\definecolor{fill-0}{RGB}{246, 249, 255} 
\definecolor{fill-1}{RGB}{255, 246, 238}
\definecolor{fill-2}{RGB}{244, 249, 241}

\definecolor{draw-0}{RGB}{54, 110, 210}
\definecolor{draw-1}{RGB}{237, 125, 49}
\definecolor{draw-2}{RGB}{112, 173, 71}

\definecolor{fill-leaf}{RGB}{248, 248, 248}
\definecolor{draw-leaf}{RGB}{135, 135, 135}

\definecolor{color1}{RGB}{192, 0, 0}        
\definecolor{color2}{RGB}{0, 112, 192}      
\definecolor{color3}{RGB}{237, 125, 49}     
\definecolor{color4}{RGB}{0, 146, 70}       
\definecolor{color5}{RGB}{128, 0, 64}       
\definecolor{color6}{RGB}{109, 120, 40}     
\definecolor{color7}{RGB}{112, 48, 160}     
\definecolor{color8}{RGB}{0, 132, 132}      

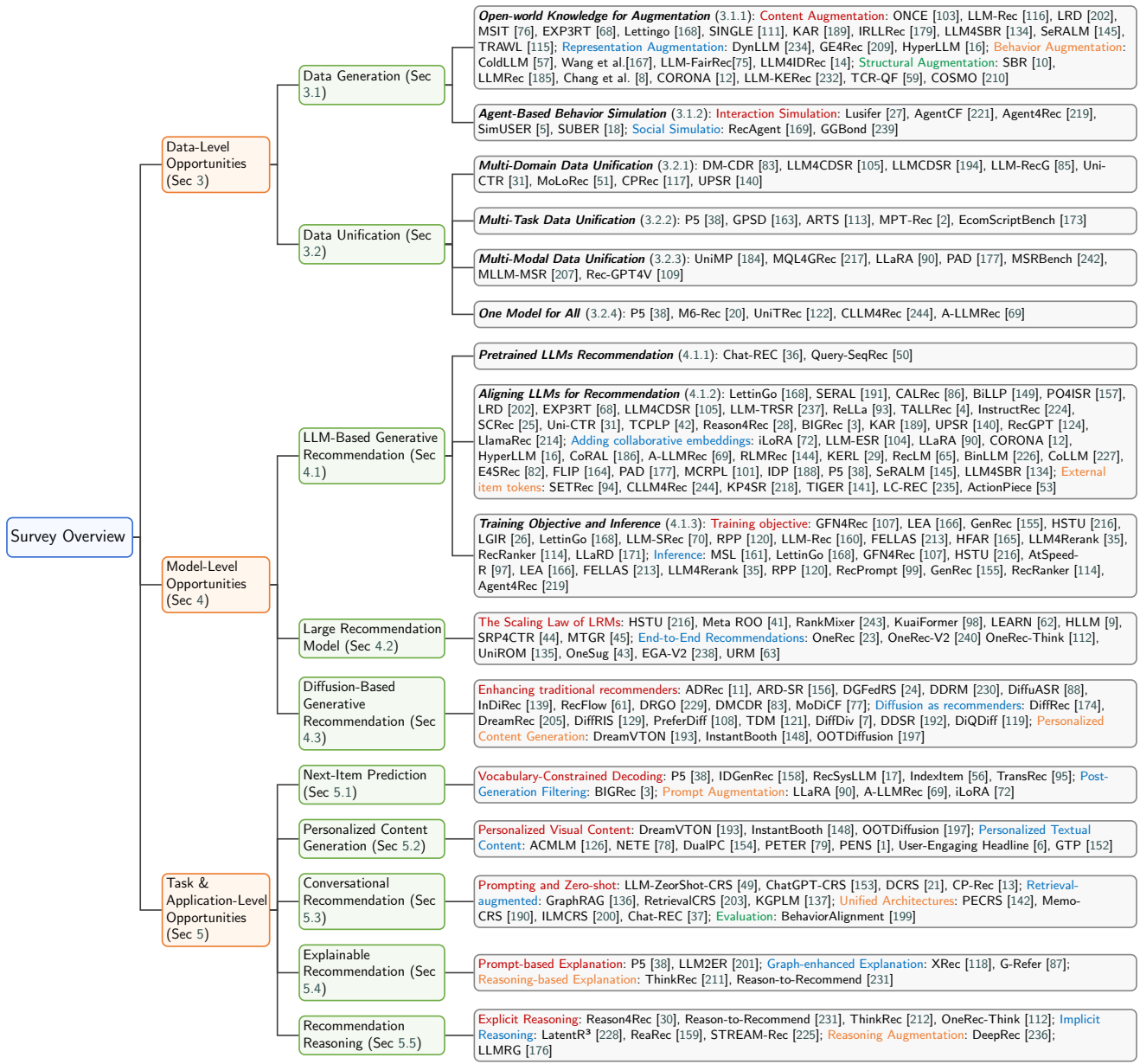
\begin{figure*}[!t]
    \centering
    \begin{adjustbox}{width=0.99\textwidth}
    \begin{forest}
        forked edges,
        for tree={
            grow=east,          
            reversed=true,      
            anchor=center,      
            parent anchor=east, 
            child anchor=west,  
            base=center,  
            ver/.style={rotate=90, child anchor=north, parent anchor=south, anchor=center},
            rectangle,          
            rounded corners,    
            fill=,              
            draw=,              
            line width=1pt,     
            minimum width=4em,  
            minimum height=2em, 
            text width=,        
            inner xsep=0.3em,   
            inner ysep=0.3em,   
            edge+={darkgray, line width=1pt},   
            s sep=10pt,         
            l sep=20pt,         
            align=left,          
            font=,              
        },
        where level=0{fill=fill-0, draw=draw-0, text width=9em, align=center, font=\large, minimum height=3em,}{},
        where level=1{fill=fill-1, draw=draw-1, text width=7.6em, align= left, font=\normalsize,}{},
        where level=2{fill=fill-2, draw=draw-2, text width=10.5em, align=left, font=\normalsize,}{},
        where level=3{fill=fill-leaf, draw=draw-leaf, text width=50em, font=\small,}{},
        [
        Survey Overview 
            [
            \parbox{7.6em}{Data-Level Opportunities (Sec \ref{sec3:Data-Level Opportunities})}
                [
                \parbox{7em}{Data Generation (Sec \ref{sec3.1:Data Generation})}
                    [
                    \parbox{50em}{
                    \textbf{\textit{Open-world Knowledge for Augmentation}} (\ref{sec3.1.1:Open-world Knowledge for Augmentation}): 
                    {\color{color1} Content Augmentation}: ONCE~\cite{ONCE-WSDM-2024}, LLM-Rec~\cite{LLM-REC-NAACL-2024}, LRD~\cite{LRD-SIGIR-2024}, MSIT~\cite{MSIT-ACL-2025}, EXP3RT~\cite{EXP3RT-SIGIR-2025}, Lettingo~\cite{LettinGo-KDD-2025}, SINGLE~\cite{SINGLE-WebConf-2024}, KAR~\cite{KAR-RecSys-2024}, IRLLRec~\cite{IRLLRec-SIGIR-2025}, LLM4SBR~\cite{LLM4SBR-TOIS-2025}, SeRALM~\cite{SeRALM-SIGIR-2024}, TRAWL~\cite{TRAWL-2024}; 
                    {\color{color2} Representation Augmentation}: DynLLM~\cite{DynLLM-2024}, GE4Rec~\cite{yinfeature},  HyperLLM~\cite{HyperLLM-SIGIR-2025};  
                    {\color{color3} Behavior Augmentation}: ColdLLM~\cite{ColdLLM-WSDM-2025}, Wang et al.\cite{wang2024large-WWW-2024}, LLM-FairRec\cite{FAIRREC-SIGIR-2025}, LLM4IDRec~\cite{LLM4IDRec-TOIS-2025};              
                    {\color{color4} Structural Augmentation}: SBR~\cite{SBR-SIGIR-2025}, LLMRec~\cite{wei2024llmrec}, Chang et al.~\cite{chang2025llms}, CORONA~\cite{CORONA-SIGIR-2025}, LLM-KERec~\cite{LLM-KERec-CIKM-2024}, TCR-QF~\cite{huang2025mitigate}, COSMO~\cite{COSMO-SIGMOD-2024}
                    }
                    ]
                    [
                    \parbox{50em}{
                    \textbf{\textit{Agent-Based Behavior Simulation}} (\ref{sec3.1.2:Agent-Based Behavior Simulation}): 
                    {\color{color1} Interaction Simulation}: Lusifer~\cite{ebrat2024lusifer}, AgentCF~\cite{AgentCF-WWW-2024}, Agent4Rec~\cite{Agent4Rec-SIGIR-2024}, STEAM~\cite{STEAM-ARXIV-2026}, SimUSER~\cite{bougie2025simuser}, SUBER~\cite{corecco2024suber}; 
                    {\color{color2} Social Simulation}: RecAgent~\cite{RecAgent-TOIS-2025}, GGBond~\cite{zhong2025ggbond}            
                    }
                    ]
                ]
                [
                \parbox{10.5em}{Data Unification \\(Sec \ref{sec3.2:Data Unification})}
                    [
                    \parbox{50em}{
                    \textbf{\textit{Multi-Domain Data Unification}} (\ref{sec3.2.1:Multi-Domain Data Unification}):
                    DM-CDR~\cite{DMCDR-KDD-2025}, LLM4CDSR~\cite{LLM4CDSR-SIGIR-2025}, LLMCDSR~\cite{LLMCDSR-TOIS-2025}, LLM-RecG~\cite{LLM-RecG-RecSys-2025}, Uni-CTR~\cite{Uni-CTR-TOIS-2025}, CPRec~\cite{ma2025large}, UPSR~\cite{UPSR-TOIS-2025}
                    }
                    ]
                    [
                    \parbox{50em}{
                    \textbf{\textit{Multi-Task Data Unification}} (\ref{sec3.2.2:Multi-Task Data Unification}):               
                    P5~\cite{P5-RecSys-2022}, GPSD~\cite{GPSD-KDD-2025}, ARTS~\cite{ARTS-TOIS-2025}, MPT-Rec~\cite{MPT-Rec-TOIS-2025}, EcomScriptBench~\cite{ECOMSCRIPT-ACL-2025}
                    }
                    ]
                    [
                    \parbox{50em}{
                    \textbf{\textit{Multi-Modal Data Unification}} (\ref{sec3.2.3:Multi-Modal Data Unification}):              
                    UniMP~\cite{UniMP-ICLR-2024}, MQL4GRec~\cite{MQL4GRec-ICLR-2025}, LLaRA~\cite{LLaRA-SIGIR-2024}, PAD~\cite{PAD-SIGIR-2025}, MSRBench~\cite{MSRBench-WWW-2025}, MLLM-MSR~\cite{harnessing-AAAI-2025}, Rec-GPT4V~\cite{Rec-GPT4V-2024} 
                    }
                    ]
                    [
                    \parbox{50em}{
                    \textbf{\textit{One Model for All}} (\ref{sec3.2.4:One Model for All}): 
                    P5~\cite{P5-RecSys-2022}, M6-Rec~\cite{M6-Rec}, UniTRec~\cite{UniTRec-ACL-2023}, CLLM4Rec~\cite{CLLM4Rec-WWW-2024}, A-LLMRec~\cite{A-LLMRec-KDD-2024}, RecCocktail~\cite{Hou_Bai_Wu_Liu_Zhang_Liu_Hong_Tang_Wang_2026}, WeaveRec~\cite{weaverec_WWW_2026}
                    }
                    ]
                ]
            ]
            [
            \parbox{7.6em}{Model-Level Opportunities (Sec \ref{sec4:Model-Level Opportunities})}
                [
                \parbox{10.5em}{LLM-Based Generative Recommendation \\(Sec \ref{sec4.1:LLM-Based Generative Recommendation})}
                    [
                    \parbox{50em}{
                    \textbf{\textit{Pretrained LLMs Recommendation}} (\ref{sec4.1.1:Pretrained LLMs Recommendation}):
                    Chat-REC~\cite{Chat-REC-2023}, Query-SeqRec~\cite{Query-SeqRec-CIKM-2022}
                    }
                    ]
                    [
                    \parbox{50em}{
                    \textbf{\textit{Aligning LLMs for Recommendation}} (\ref{sec4.1.2:Aligning LLMs for Recommendation}):                   
                    LettinGo~\cite{LettinGo-KDD-2025}, SERAL~\cite{SERAL-KDD-2025}, CALRec~\cite{CALRec-RecSys-2024}, BiLLP~\cite{BiLLP-SIGIR-2024}, PO4ISR~\cite{PO4ISR-SIGIR-2024}, LRD~\cite{LRD-SIGIR-2024}, EXP3RT~\cite{EXP3RT-SIGIR-2025}, LLM4CDSR~\cite{LLM4CDSR-SIGIR-2025}, LLM-TRSR~\cite{LLM-TRSR-WWW-2024}, ReLLa~\cite{ReLLa-WWW-2024}, TALLRec~\cite{TALLRec-RecSys-2023}, InstructRec~\cite{InstructRec-TOIS-2025}, SCRec~\cite{SCRec-TOIS-2025}, Uni-CTR~\cite{Uni-CTR-TOIS-2025}, TCPLP~\cite{TCPLP-TOIS-2025}, Reason4Rec~\cite{Reason4Rec-Arxiv-2025}, BIGRec~\cite{BIGRec-TORS-2025}, KAR~\cite{KAR-RecSys-2024}, UPSR~\cite{UPSR-TOIS-2025}, RecGPT~\cite{RecGPT-ACL-2024}, LlamaRec~\cite{Llamarec-2023};
                    {\color{color2} Adding collaborative embeddings}:
                    iLoRA~\cite{iLoRA-NeurIPS-2024}, LLM-ESR~\cite{LLM-ESR-NeurIPS-2024}, LLaRA~\cite{LLaRA-SIGIR-2024}, CORONA~\cite{CORONA-SIGIR-2025}, HyperLLM~\cite{HyperLLM-SIGIR-2025}, CoRAL~\cite{CoRAL-KDD-2024}, A-LLMRec~\cite{A-LLMRec-KDD-2024}, RLMRec~\cite{RLMRec-WWW-2024}, RecLM~\cite{RecLM-ACL-2025}, BinLLM~\cite{BinLLM-ACL-2024}, CoLLM~\cite{CoLLM-TKDE-2025}, E4SRec~\cite{E4SRec-WWW-2024}, FLIP~\cite{FLIP-RecSys-2024}, PAD~\cite{PAD-SIGIR-2025}, MCRPL~\cite{MCRPL-TOIS-2024}, IDP~\cite{IDP-TOIS-2025}, P5~\cite{P5-RecSys-2022}, SeRALM~\cite{SeRALM-SIGIR-2024}, LLM4SBR~\cite{LLM4SBR-TOIS-2025};
                    {\color{color3} External item tokens}:
                    SETRec~\cite{SETRec-SIGIR-2025}, CLLM4Rec~\cite{CLLM4Rec-WWW-2024}, KP4SR~\cite{KP4SR-MM-2023}, TIGER~\cite{TIGER-NeurIPS-2023}, LC-REC~\cite{LC-Rec-ICDE-2024}, ActionPiece~\cite{ActionPiece-ICML-2025}
                    }
                    ]
                    [
                    \parbox{50em}{
                    \textbf{\textit{Training Objective and Inference}} (\ref{sec4.1.3:Training Objective and Inference}):
                    {\color{color1} Training objective}:
                    GFN4Rec~\cite{GFN4Rec-KDD-2023}, LEA~\cite{LEA-SIGIR-2024}, GenRec~\cite{GenRec-RecSys-2023}, HSTU~\cite{HSTU-ICML-2024}, LGIR~\cite{LGIR-AAAI-2024}, LettinGo~\cite{LettinGo-KDD-2025}, LLM-SRec~\cite{LLM-SRec-KDD-2025}, RPP~\cite{RPP-TOIS-2025}, LLM-Rec~\cite{LLM-Rec-TOIS-2025}, FELLAS~\cite{FELLAS-TOIS-2024}, HFAR~\cite{HFAR-TOIS-2025}, LLM4Rerank~\cite{LLM4Rerank-WWW-2025}, RecRanker~\cite{RecRanker-TOIS-2025}, LLaRD~\cite{LLaRD-WWW-2025};
                    {\color{color2} Inference}:
                    MSL~\cite{MSL-SIGIR-2025}, LettinGo~\cite{LettinGo-KDD-2025}, GFN4Rec~\cite{GFN4Rec-KDD-2023}, HSTU~\cite{HSTU-ICML-2024}, AtSpeed-R~\cite{AtSpeed-ICLR-2025}, LEA~\cite{LEA-SIGIR-2024}, FELLAS~\cite{FELLAS-TOIS-2024}, LLM4Rerank~\cite{LLM4Rerank-WWW-2025}, RPP~\cite{RPP-TOIS-2025}, RecPrompt~\cite{RecPrompt-CIKM-2024}, GenRec~\cite{GenRec-RecSys-2023}, RecRanker~\cite{RecRanker-TOIS-2025}, Agent4Rec~\cite{Agent4Rec-SIGIR-2024}
                    }
                    ]
                ]
                [
                \parbox{10.5em}{Large Recommendation Model (Sec \ref{sec4.2:Large Recommendation Model})}
                    [
                    \parbox{50em}{
                    {\color{color1} The Scaling Law of LRMs}:
                    HSTU~\cite{HSTU-ICML-2024},
                    Meta ROO~\cite{Meta-ROO-Arxiv-2025},
                    RankMixer~\cite{RankMixer-Arxiv-2025},
                    KuaiFormer~\cite{liu2024-kuaiformer}, LEARN~\cite{LEARN-AAAI-2025}, HLLM~\cite{chen-HLLM-2024}, SRP4CTR~\cite{SRP4CTR-CIKM-2024}, MTGR~\cite{MTGR-arxiv-2025};
                    {\color{color2} End-to-End Recommendations}:
                    OneRec~\cite{OneRec-2025},
                    OneRec-V2~\cite{OneRec-v2-arxiv-2025}
                    OneRec-Think~\cite{OneRec-Think-Arxiv-2025},
                    UniROM~\cite{Qiuone-2025}, OneSug~\cite{guo-OneSug-2025}, EGA-V2~\cite{zheng-EGAV2-2025}, URM~\cite{URM-2025}
                    }
                    ]
                ]
                [
                \parbox{10.5em}{Diffusion-Based Generative Recommendation \\(Sec \ref{sec4.3:Diffusion-Based Generative Recommendation})}
                    [
                    \parbox{50em}{                 
                    {\color{color1} Enhancing traditional recommenders}:
                    ADRec~\cite{ADRec-KDD-2025}, ARD-SR~\cite{ARD-SR-WWW-2025}, DGFedRS~\cite{DGFedRS-TOIS-2025}, DDRM~\cite{DDRM-SIGIR-2024}, DiffuASR~\cite{Diffurec-TOIS-2025}, InDiRec~\cite{InDiRec-SIGIR-2025}, RecFlow~\cite{RecFlow-ICML-2025}, DRGO~\cite{DRGO-WWW-2025}, DMCDR~\cite{DMCDR-KDD-2025}, MoDiCF~\cite{MoDiCF-WWW-2025};                    
                    {\color{color2} Diffusion as recommenders}:
                    DiffRec~\cite{DiffRec-SIGIR-2023}, DreamRec~\cite{DreamRec-NeurIPS-2023}, DiffRIS~\cite{DiffRIS-WWW-2024}, PreferDiff~\cite{PreferDiff-ICLR-2025}, TDM~\cite{TDM-SIGIR-2025}, DiffDiv~\cite{DiffDiv-SIGIR-2025}, DDSR~\cite{DDSR-NeurIPS-2024}, DiQDiff~\cite{DiQDiff-WWW-2025}, HorizonRec~\cite{zha2026align};
                    {\color{color3} Personalized Content Generation}:
                    DreamVTON~\cite{Dreamvton-MM-2024}, InstantBooth~\cite{Instantbooth-CVPR-2024}, OOTDiffusion~\cite{Ootdiffusion-AAAI-2025}
                    }
                    ]
                ]
            ]
            [
            \parbox{7.6em}{Task \& Application-Level Opportunities (Sec \ref{sec5:Task and Application-Level Opportunities})}
                [
                \parbox{10.5em}{Next-Item Prediction (Sec \ref{sec5.1:Next-Item Prediction})}
                    [
                    \parbox{50em}{
                    {\color{color1} Vocabulary-Constrained Decoding}:
                    P5~\cite{P5-RecSys-2022}, IDGenRec~\cite{IDGenRec-SIGIR-2024}, RecSysLLM~\cite{RecSysLLM-arxiv-2023}, IndexItem~\cite{IndexItem-SIGIRAP-2023}, TransRec~\cite{TransRec-KDD-2024};
                    {\color{color2} Post-Generation Filtering}:
                    BIGRec~\cite{BIGRec-TORS-2025};
                    {\color{color3} Prompt Augmentation}:
                    LLaRA~\cite{LLaRA-SIGIR-2024}, A-LLMRec~\cite{A-LLMRec-KDD-2024}, iLoRA~\cite{iLoRA-NeurIPS-2024}
                    }
                    ]
                ]
                [
    \parbox{10.5em}{Personalized Content Generation (Sec \ref{sec5.2:Recommended Item Generation})}
        [
        \parbox{50em}{
        {\color{color1} Personalized Visual Content}:
        DreamVTON~\cite{Dreamvton-MM-2024}, InstantBooth~\cite{Instantbooth-CVPR-2024}, OOTDiffusion~\cite{Ootdiffusion-AAAI-2025};
        {\color{color2} Personalized Textual Content}:
        ACMLM~\cite{Justifying-EMNLPIJCNLP-2019}, NETE~\cite{TemplateExplanation-CIKM-2020}, DualPC~\cite{DualLearning-WWW-2020}, PETER~\cite{PETER-ACL-2021}, PENS~\cite{PENS-ACL-2021}, User-Engaging Headline~\cite{Generating-ACL-2023}, GTP~\cite{General-TACL-2023}
        }
        ]
    ]
    [
    \parbox{10.5em}{Conversational Recommendation \\(Sec \ref{sec5.3:Conversational Recommendation})}
        [
        \parbox{50em}{
        {\color{color1} Prompting and Zero-shot}:
        LLM-ZeorShot-CRS~\cite{zeroshotCRS-CIKM-2023}, ChatGPT-CRS~\cite{chatgptCRS-Arxiv-2024}, DCRS~\cite{DCRS-SIGIR-2024}, CP-Rec~\cite{CP-Rec-AAAI-2023};
        {\color{color2} Retrieval-augmented}:
        GraphRAG~\cite{GraphRAGCRS-PAKDD-2025}, RetrievalCRS~\cite{RetrievalCRS-RecSys-2024}, KGPLM~\cite{KGPLM-CRS-TNNLS-2025};
        {\color{color3} Unified Architectures}:
        PECRS~\cite{ravaut-etal-2024-parameter}, MemoCRS~\cite{MemoCRS-CIKM-2024}, ILMCRS~\cite{ILMCRS-Arxiv-2025}, Chat-REC~\cite{Chat-REC-2023};
        {\color{color4} Evaluation}:
        BehaviorAlignment~\cite{BehaviorAlignment-SIGIR-2024}
        }
        ]
    ]
    [
    \parbox{10.5em}{Explainable Recommendation \\(Sec \ref{sec5.4:Explainable Recommendation})}
        [
        \parbox{50em}{
        {\color{color1} Prompt-based Explanation}:
        P5~\cite{P5-RecSys-2022}, LLM2ER~\cite{LLM2ER-AAAI-2024};
        {\color{color2} Graph-enhanced Explanation}:
        XRec~\cite{XRec-EMNLP-2024}, G-Refer~\cite{G-Rec-WWW-2025};
        {\color{color3} Reasoning-based Explanation}:
        ThinkRec~\cite{ThinkRec-Arxiv-2025}, Reason-to-Recommend~\cite{Reason-to-Recommend-Arxiv-2025}
        }
        ]
    ]
    [
    \parbox{10.5em}{Recommendation Reasoning (Sec \ref{sec5.5:Recommendation Reasoning})}
        [
        \parbox{50em}{
        {\color{color1} Explicit Reasoning}:
        Reason4Rec~\cite{Reason4Rec-Arxiv-2025}, Reason-to-Recommend~\cite{Reason-to-Recommend-Arxiv-2025}, ThinkRec~\cite{ThinkRec-Arxiv-2025}, OneRec-Think~\cite{OneRec-Think-Arxiv-2025};
        {\color{color2} Implicit Reasoning}:
         LatentR³~\cite{LatantR3-Arxiv-2025}, ReaRec~\cite{ReaRec-Arxiv-2025}, STREAM-Rec~\cite{Stream-Rec-Arxiv-2025};
        {\color{color3} Reasoning Augmentation}:
        DeepRec~\cite{DeepRec-Arxiv-2025};
        LLMRG~\cite{LLMRG-AAAI-2024}
        }
        ]
    ]
]
]
    \end{forest}
    \end{adjustbox}
    \caption{Taxonomy of research on generative recommendation.}
\end{figure*}

%% file: opportunities.tex
\FloatBarrier
\section{Preliminary and Background}
\label{sec2:Foundations of Generative Recommendation}

In this section, we begin with a foundational overview of traditional discriminative and generative recommendation models. We then explore the potential benefits that generative models can bring to recommender systems.

\subsection{Preliminaries of Discriminative \& Generative Recommendation}
\label{sec2.1:Preliminaries of Discriminative and Generative Recommendation}
Strictly speaking, discriminative models learn the conditional probability $P(y|x)$, or directly map input $x$ to predict output $y$. 
In contrast, generative models learns the joint probability distribution $P(x,y)$, meaning it models how both the input $x$ and the label $y$ are generated together. 
\subsubsection{Discriminative Recommendation Models}
\label{sec2.1.1:Discriminative Recommendation Models}
For the recommendation task, discriminative recommendation models focus on learning a scoring or ranking function $f(u,i)$ that estimates the relevance or affinity between a user $u$ and an item $i$. According to the recommendation system construction pipeline, we can divide it into three parts: data, model, and task.

\textbf{Data Preparation.} Given training data consisting of tuples $\mathcal{D}=\{(u, i, y_{ui})\}$, where $u$ represents a user, $i$ represents an item, and $y_{ui}$ denotes the observed interaction (ratings for explicit feedback or binary values $\{0,1\}$ for implicit feedback). The inputs $u$ and $i$ can be one-hot IDs in collaborative filtering methods. Auxiliary content data are commonly utilized to enrich $u$ and $i$, including but not limited to user social networks and profiles for users, and multimedia descriptions (images, videos, texts, audio) and knowledge graphs for items.

\textbf{Model Construction.} At the training time, discriminative recommendation methods start by utilizing embedding layers to map each user and item to a dense embedding vector: $\mathbf{e}_u=\phi_u(u), \mathbf{e}_i=\phi_i(i)$.
$\phi_u(u)$ and $\phi_i(i)$ are embedding layers for users and items, often simple lookup tables, but can include Multi-Layer Perceptrons (MLPs), Graph Neural Networks (GNNs), Convolutional Neural Networks (CNNs), transformers, etc.
Then, discriminative recommendation models computes a matching score between the user and item embeddings: $f_{ui}=\operatorname{Score}\!\left(\mathbf{e}_u, \mathbf{e}_i\right)$.
Common scoring functions include inner product, distance-based metrics, and neural network-based metrics (e.g., MLP and CNN).
The discriminative recommendation models are usually trained to discriminate between positive and negative interactions.
Commonly used loss functions include Mean Squared Error~(MSE) loss $\mathcal{L}_\text{rating}$ and Binary Cross Entropy~(BCE) loss $\mathcal{L}_\text{point}$ for explicit feedback, and Bayesian Personalized Ranking~(BPR) loss $\mathcal{L}_{\text{pair}}$ for implicit feedback. 
\begin{flalign}
&\mathcal{L}_{\text{rating}}=\frac{1}{N} \sum_{i=1}^N\left(y_{u i}-f(u,i)\right)^2, &&\\[-2pt]
&\mathcal{L}_{\text{point}} = -\sum_{(u,i)\in\mathcal{D}} [y_{ui}\log\sigma(f_{ui}) + (1-y_{ui})\log(1-\sigma(f_{ui}))], &&\\[-2pt]
&\mathcal{L}_{\text{pair}} = - \sum_{(u,i^+,i^-) \in D}\log\sigma(f_{ui^+}-f_{ui^-}). &&
\end{flalign}

\textbf{Recommendation Task.} For discriminative recommendation systems, the final recommendation task is mostly to select the top k items that users might like from a candidate list, which we refer to as top-k recommendation. At inference time, given a user $u$ and candidate item set $\mathcal{I}$, the discriminative recommendation models compute scores and ranks the items.
\begin{flalign}
&\hat{i}=\arg\max_{i \in \mathcal{I}} f(u, i),\qquad
\operatorname{TopK}_u=\operatorname{Top\text{-}K}_{i \in \mathcal{I}} f(u, i). &&
\end{flalign}
This process requires calculating the user's matching score for each item in the candidate list, then ranking and selecting the top-k items for recommendation.

\subsubsection{Generative Recommendation Models}
\label{sec2.1.2:Generative Recommendation Models}
In this survey, we systematically categorize all recommendation approaches that utilize generative models within the overarching paradigm of generative recommendation. We define generative recommendation as a broad recommendation paradigm in which generative models~(such as LLMs, diffusion models) are utilized in stages of the recommendation pipeline.
We can still categorize the recommendation pipeline into three major phases: data, model, and tasks. We identify three paradigms of generative recommendation in current research: data-level synthesis, model-level recommendation, and task-level generation.

\textbf{Data-Level Synthesis.} 
Generative models are used to synthesize training data, including both user/item features and interaction records. This is especially useful in scenarios like cold-start problems or sparsity in the dataset. Formally, let $\mathcal{D}$ represents the original dataset consisting of user set $\mathcal{U}$, item set $\mathcal{I}$, and interaction set $\mathcal{Y}$, generative models produce:
$$\mathcal{U}', \mathcal{I}', \mathcal{Y}' = G_{\text{data}}(\mathcal{U}, \mathcal{I}, \mathcal{Y} | \theta_g),$$
where $G_{\text{data}}$ generates synthetic user features, item features, and interaction records to create an enriched training dataset for recommendation models.

\textbf{Model-Level Recommendation.} 
At the model level, generative models serve as the core recommendation engine to directly learn user preferences and generate personalized recommendations. Current mainstream work in this paradigm can be categorized into three major approaches: LLM-based approaches, large recommendation models, and diffusion model-based approaches. 
(i) LLM-based approaches leverage pre-trained language models as recommendation backbones. These methods convert recommendation data into samples with textual input and output, and model the recommendation task as the neural language generation process. (ii) Large recommendation models scale up traditional recommendation architectures with generative components, employing massive parameters to model complex user-item interactions and generate high-quality recommendations. (iii) Diffusion model-based approaches treat recommendation as a denoising process, learning to generate user preferences or item rankings through iterative refinement from noise to meaningful recommendation signals.

\textbf{Task-Level Generation.} 
At the task level, generative models reformulate recommendation as a generation task, producing recommendation outputs in natural language or structured formats. This paradigm not only addresses traditional recommendation tasks such as sequential recommendation and click-through rate prediction through generative approaches, but also enables novel recommendation capabilities including generating personalized explanations for recommendations, creating conversational recommendation dialogues, producing item reviews and descriptions, creating virtual items, and creating multi-modal recommendation content that combines text, images, and other media formats. This paradigm opens up new possibilities for interpretable and interactive recommendation experiences.

\subsection{Advantages of Generative Models}
\label{sec2.2:Benefits of Generative Models}
In this subsection, we summarize the key advantages that generative models bring to recommender systems by addressing fundamental challenges in understanding users and items.

\textbf{World Knowledge Integration.}
Effective recommendations require understanding not only user-item interaction patterns but also the rich semantic information and real-world context of items themselves. Traditionally, incorporating world knowledge into recommendations required explicit content enrichment methods, such as extracting knowledge graphs and leveraging auxiliary content information. These approaches often involved multi-stage pipelines with separate modules for knowledge extraction and integration. Generative models, particularly LLMs pre-trained on vast and diverse datasets, inherently encode extensive world knowledge about entities, events, relationships, and cultural contexts. By adopting a generative paradigm for recommendation, systems can directly leverage this embedded knowledge without requiring step-by-step knowledge extraction pipelines. For example, when recommending movies, a generative model can naturally incorporate information about directors, actors, cultural trends, or even recent events, all without explicit knowledge base construction. This seamless integration of world knowledge enables more contextually aware and informed recommendations.

\textbf{Natural Language Understanding.}
Personalized recommendation fundamentally relies on understanding users. While implicit behavioral signals (clicks, purchases, ratings) have been the primary data source, users naturally express their preferences, needs, and intentions through language, including search queries, reviews, conversations, and feedback. Traditional recommendation systems often struggle to effectively process and understand these rich natural language signals. Generative models, particularly LLMs, with their advanced natural language understanding capabilities, can directly interpret user expressions in free-form text, understanding nuances, context, and intent. This enables recommendation systems to capture user preferences expressed in natural language, support conversational recommendation interfaces, and understand complex, multi-faceted queries. For instance, when a user asks "I want something relaxing but not boring for a Friday night," a generative model can parse this nuanced request and generate appropriate recommendations, bridging the gap between how users naturally communicate and how recommendation systems operate.

\textbf{Reasoning Capabilities.}
User decision-making in recommendation scenarios often involves more than simple preference matching; it requires reasoning. Particularly in complex decision contexts (such as selecting financial products, planning travel, or making healthcare choices), users engage in multi-step reasoning processes that consider trade-offs, constraints, and causal relationships. Traditional recommendation models typically rely on pattern matching and correlation, lacking explicit reasoning abilities. Generative models, with their emergent reasoning capabilities, can model the logical processes behind user decisions. They can understand "why" a user might prefer one item over another by considering feature relationships, temporal sequences, and contextual factors. This reasoning ability allows generative recommendation systems to provide not just item suggestions but also explanations and justifications, making recommendations more transparent and trustworthy.

\textbf{Scaling Law.}
The scaling law observed in large language models demonstrates that model performance improves significantly as both model size and training data increase. This principle offers a promising pathway for building more capable recommendation systems. By scaling up generative models with larger parameters and more diverse training data, recommendation systems can capture deeper patterns in user behavior and item characteristics. The scaling law suggests that as we increase the scale of generative models for recommendation, they can exhibit emergent capabilities that were not explicitly programmed, such as better understanding of user intent, more nuanced preference modeling, and improved handling of complex recommendation scenarios. This provides a systematic approach to advancing recommendation quality: instead of manually engineering features or designing intricate model architectures for each specific scenario, we can leverage the scaling law to achieve better performance through increased scale. Furthermore, larger-scale generative models trained on massive amounts of recommendation data across multiple platforms and contexts can learn universal patterns in user-item interactions, enabling more sophisticated and accurate personalized recommendations.

\textbf{Generative Capabilities for Novel Recommendations.}
Unlike discriminative recommendation models that rank or select from existing item candidates, generative models can create novel content and recommendations. This capability is particularly valuable in cold-start scenarios, where new users or new items lack historical interaction data. Generative models can synthesize recommendations by drawing on broader patterns, user archetypes, and item similarities, ensuring meaningful suggestions even without explicit past behavior. Moreover, they can generate diverse and creative recommendations that break free from the filter bubble problem common in traditional systems. For instance, instead of merely ranking a fixed set of products, a generative model could suggest customized bundles, personalized content variations, or even generate entirely new item descriptions tailored to individual user preferences.

\subsection{When Does Generative Recommendation Outperform Discriminative Approaches?}
While the advantages outlined above highlight the transformative potential of generative models, it is important to recognize that their superiority over discriminative approaches is not universal. It depends critically on the task, data, and deployment context. Based on existing empirical evidence, we identify the following three main conditions under which generative recommendation demonstrates genuine and reliable effectiveness gains.
(1) Data-sparse and cross-domain scenarios. When interaction data is scarce or distributed across domains, LLMs' embedded world knowledge compensates for insufficient behavioral signals, leading to consistent improvements in cold-start and zero-shot settings that discriminative models struggle to match.
(2) Inherently generative tasks. For tasks such as conversational recommendation, explainable recommendation, and personalized content creation, generative models hold a fundamental advantage, as these tasks require open-ended generation capabilities that discriminative scoring functions cannot provide.
(3) Large-scale training regimes. As demonstrated by HSTU~\cite{HSTU-ICML-2024} and related large recommendation models, generative architectures benefit from scaling laws that allow performance to improve predictably with increased model size and data, a property that discriminative models plateau on beyond a certain scale.

These observations suggest that the effectiveness of generative recommendation is deeply intertwined with how generative models are leveraged across the entire recommendation pipeline, from enriching sparse data and scaling model capacity to enabling novel task formulations, which motivates the data-model-task framework adopted in this survey.

%% file: data_lyx.tex
\section{Data-Level Opportunities}
\label{sec3:Data-Level Opportunities}
\begin{figure*}[!t]
    \centering \includegraphics[width=0.98\textwidth]{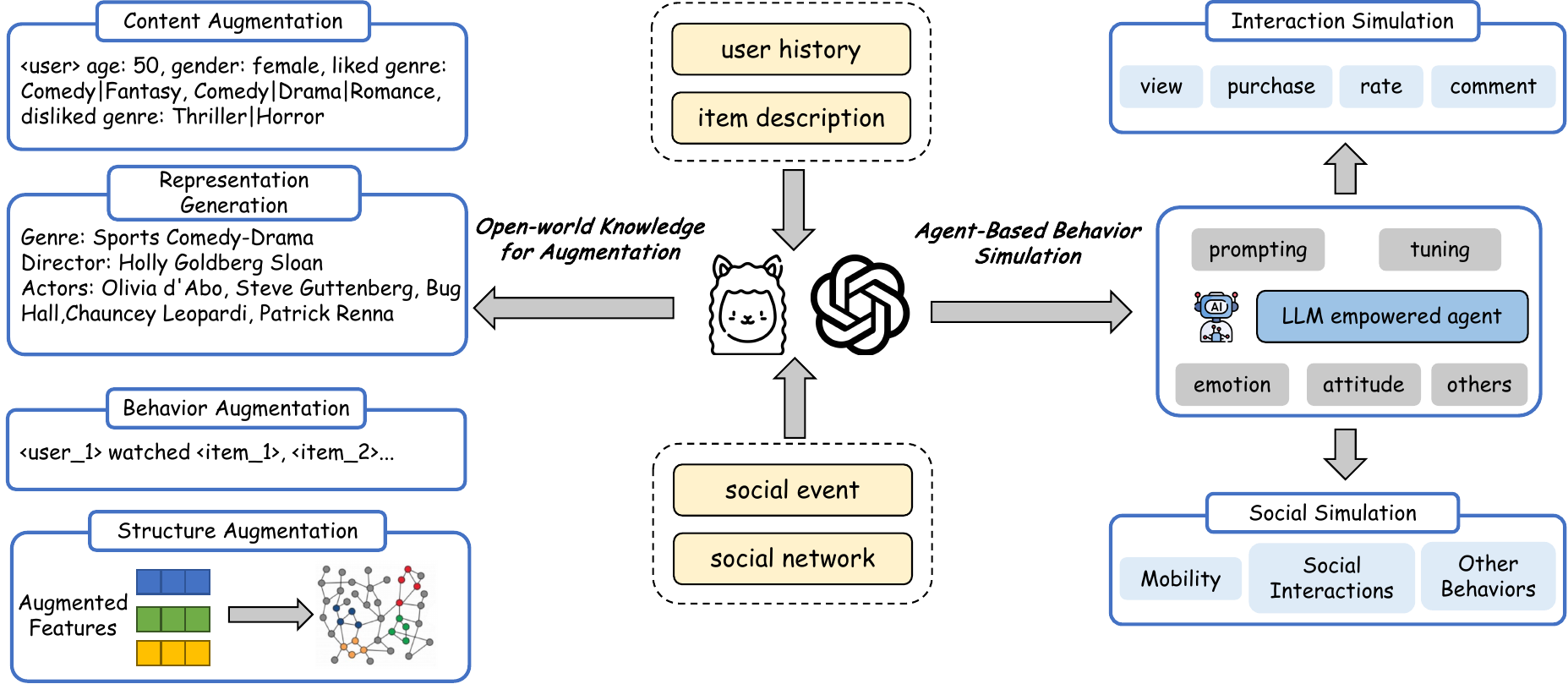}
    \caption{Outline of key techniques in LLM-empowered data generation.}
    \label{fig:data_stage_1}
\end{figure*}

LLMs have unlocked new possibilities for data-centric advancements in recommender systems by enabling effective data generation and unification. Unlike traditional methods that depend solely on existing datasets, LLMs empower systems to actively generate rich user and item data, as well as simulate realistic interaction scenarios. Additionally, LLMs facilitate the unification of heterogeneous data sources across tasks, domains, and modalities, addressing longstanding challenges of data sparsity and fragmentation. This section reviews recent progress in two primary areas: (1) \textbf{data generation}, encompassing open-world knowledge augmentation and agent-based behavior simulation, as shown in Fig. \ref{fig:data_stage_1}, and (2) \textbf{data unification}, which integrates diverse data types to form coherent inputs for recommendation models. Together, these advances form a foundational basis for building more adaptive, generalizable, and generative recommendation systems.

\begin{table*}[t]
\centering
\caption{Open-world Knowledge Data Augmentation for Recommender Systems.}
\label{tab:data-gugmentation}
\begin{tabular}{p{3.2 cm} p{6.4cm} p{6.2cm}}
\toprule
\textbf{Category} & \textbf{Representative Works} & \textbf{Description / Focus} \\
\midrule
\textbf{Content Augmentation} & 
ONCE (WSDM’24), LLM-Rec (NAACL’24), LRD (SIGIR’24), MSIT (ACL’25), EXP3RT (SIGIR’25), Lettingo (KDD’25), SINGLE (WWW’24), KAR (RecSys’24), IRLLRec (SIGIR’25), LLM4SBR (TOIS’25), SeRALM (SIGIR’24), TRAWL (ArXiv’24) & 
Generate natural-language user/item profiles, summarize histories, enrich sparse metadata, and align textual semantics with feedback. \\
\midrule
\textbf{Representation Augmentation} & 
DynLLM (ArXiv’24), GE4Rec (ICML’24), HyperLLM (SIGIR’25) & 
Automated feature construction, multimodal attribute extraction, external knowledge distillation, and hierarchical category generation. \\
\midrule
\textbf{Behavior Augmentation} & 
ColdLLM (WSDM’25), Wang et al. (WWW’25), LLM-FairRec (SIGIR’25), LLM4IDRec (TOIS’25) & 
Generate synthetic user–item interactions, simulate cold-start preferences, ensure fairness, and integrate pseudo-interactions into ID-based pipelines. \\
\midrule
\textbf{Structure Augmentation} & 
SBR (SIGIR’25), LLMRec (WSDM’24), Chang et al. (AAAI’25), CORONA (SIGIR’25), LLM-KERec (CIKM’24), TCR-QF (IJCAI’25), COSMO (SIGMOD’24) & 
Relation discovery, graph completion, social network generation, subgraph retrieval, knowledge graph construction \& distillation. \\
\bottomrule
\end{tabular}
\end{table*}

\subsection{Data Generation}
\label{sec3.1:Data Generation}
\subsubsection{Open-World Knowledge for Augmentation}
\label{sec3.1.1:Open-world Knowledge for Augmentation}

LLMs unlock unprecedented opportunities at the data foundation of recommender systems by leveraging their vast open-world knowledge, natural language understanding, and generative capabilities. Moving beyond passive data consumption, LLMs actively enrich, synthesize, and unify recommendation data, driving a new wave of data-centric innovations. This section reorganizes recent advances into four intuitive dimensions of augmentation: 
\textbf{Content}, \textbf{Representation}, \textbf{Behavior}, and \textbf{Structure}, which collectively enable more adaptive, generalizable, and knowledge-aware recommendation models. In Table \ref{tab:data-gugmentation}, we summarize the four types of data augmentation methods.

\textbf{Content Augmentation.}
LLMs have been extensively used to generate natural-language representations that enrich sparse behavioral signals and construct expressive user and item profiles. 
Current recommendation performance is constrained by the inherent limitations of incomplete or insufficient information in item descriptions. The world knowledge stored within LLMs can help overcome these limitations, enabling the extraction of more informative text and enriching available training data~\cite{ONCE-WSDM-2024}. LLM-REC~\cite{LLM-REC-NAACL-2024} employs diverse prompting strategies to extract key insights from the rich world knowledge stored within LLMs, thereby enriching input texts. LRD~\cite{LRD-SIGIR-2024} utilizes variational reasoning to uncover latent inter-item relationships. MSIT~\cite{MSIT-ACL-2025} leverages multimodal large language models (MLLMs) to mine latent item attributes from images and text via self-corrective instruction tuning. Some approaches prompt large models to extract multifaceted, diverse user profiles from interaction logs and review information~\cite{EXP3RT-SIGIR-2025, LettinGo-KDD-2025}. SINGLE~\cite{SINGLE-WebConf-2024} extracts users' constant preferences and instant interests from read texts. KAR~\cite{KAR-RecSys-2024} employs decomposable prompts to parse user intent, while IRLLRec~\cite{IRLLRec-SIGIR-2025} aligns cross-modal signals to construct multimodal intent representations. LLM4SBR~\cite{LLM4SBR-TOIS-2025} integrates behavioral and semantic clues for multi-perspective intent modeling.
However, raw external knowledge remains poorly aligned with recommendation objectives. SeRALM~\cite{SeRALM-SIGIR-2024} designs alignment-targeted prompts to guide LLMs in generating item descriptions aligned with recommendation goals, filtering out irrelevant noise generated by the model. Lettingo~\cite{LettinGo-KDD-2025} employs Direct Preference Optimization (DPO) based on the impact of generated user profiles on recommendation outcomes, ensuring profile adaptability and effectiveness. TRAWL~\cite{TRAWL-2024} encodes generated textual information into embeddings and utilizes adapters to align them with the recommendation task space.

Together, these approaches demonstrate how textual augmentation transforms unstructured data into structured, interpretable, and adaptive profiles that improve recommendation quality.

\textbf{Representation Augmentation.}
Traditional feature engineering is increasingly replaced by LLM-driven generative approaches, which automate the construction of semantically rich and task-specific features. 
DynLLM~\cite{DynLLM-2024} employs an LLM as a content encoder to extract content representations, then refines and integrates the model for recommendation tasks, thereby enhancing user modeling and content understanding. HyperLLM~\cite{HyperLLM-SIGIR-2025} generates hierarchical category labels for hyperbolic embeddings.GE4Rec~\cite{yinfeature} proposes a generative “feature generation” paradigm, which predicts each feature embedding based on the concatenation of all feature embeddings.
These advances highlight LLMs’ role as automated feature generators that bridge unstructured inputs and structured model representations.

\FloatBarrier
\textbf{Behavior Augmentation.}
Data sparsity and cold-start challenges are key constraints limiting recommendation systems, and LLMs offer opportunities to address these challenges. Through appropriate prompting, LLMs can understand user behavior and generate context for content of interest to users. Thus, we can leverage the reasoning and generalization capabilities of LLMs to generate pseudo-interactions aligned with the preferences of cold-start users~\cite{wang2024large-WWW-2024, ColdLLM-WSDM-2025} and users unfairly treated due to insufficient interactions~\cite{FAIRREC-SIGIR-2025} by generating pseudo-interactions aligned with their preferences, addressing cold-start and fairness issues from a data perspective. Specifically, ColdLLM~\cite{ColdLLM-WSDM-2025} employs a coupled-funnel architecture of filters and refiners to screen cold-start users and simulate interactions. LLM-FairRec~\cite{FAIRREC-SIGIR-2025} employs fairness-aware prompts to generate fair pseudo-interactions aligned with the preferences of unfairly treated minority users in recommendation systems.
Additionally, some approaches explore the potential of LLMs in pure ID-based recommendation systems. LLM4IDRec~\cite{LLM4IDRec-TOIS-2025} leverages fine-tuned LLMs to augment interaction data in ID format, thereby enhancing traditional ID-based recommendation systems.

\textbf{Structure Augmentation.}
Beyond individual features or interactions, LLMs are increasingly leveraged to induce higher-level semantic structures~(e.g., relations and graphs), which support structured reasoning. 
At a structural level, SBR~\cite{SBR-SIGIR-2025} aligns item features with hierarchical intents, LLMRec~\cite{wei2024llmrec} infers missing nodes and edges in graphs, and Chang et al.~\cite{chang2025llms} simulate realistic social networks.
Knowledge graph construction and refinement are also explored: CORONA~\cite{CORONA-SIGIR-2025} retrieves intent-aware subgraphs, LLM-KERec~\cite{LLM-KERec-CIKM-2024} infers new triples, TCR-QF~\cite{huang2025mitigate} improves completion with query-aware generation, and COSMO~\cite{COSMO-SIGMOD-2024} distills e-commerce knowledge graphs for recommendation.
Collectively, structural augmentation enables knowledge-infused representations that enhance explainability, robustness, and generalization.

\subsubsection{Agent-Based Behavior Simulation}
\label{sec3.1.2:Agent-Based Behavior Simulation}
With recent advancements in LLMs, researchers are leveraging their three fundamental capabilities to design LLM-driven agents: (i) the ability to perceive and understand the environment, (ii) reasoning techniques that integrate task requirements with corresponding rewards to design task planning, and (iii) the generation of text resembling human language. These agents can comprehend complex user preferences, generate contextual recommendations through fine-grained dynamic simulations of user behavior and interactions, and achieve more nuanced decision-making beyond simple feature matching. Agents can simulate cognition, memory, emotions, decision-making, and reflection to generate more authentic user profiles, providing rich signals for training and evaluating recommendation systems. We divide this research direction into two sub-directions: interaction simulation and social simulation.

\textbf{Interaction Simulation.}
With recent advancements in LLMs, LLM-based agents have demonstrated impressive capabilities in autonomous interaction and decision-making. Applications in the recommendation domain that utilize agents to simulate users~\cite{ebrat2024lusifer} typically combine memory modules, role modeling, and reflective loops to generate adaptive and interpretable user behavior simulations.
Specifically, Agent4Rec~\cite{Agent4Rec-SIGIR-2024} scales this approach to user agents with factual and emotional memories, simulating diverse behaviors, enabling causal behavior analysis. AgentCF~\cite{AgentCF-WWW-2024} simultaneously simulates user agents and item agents, modeling the collaborative filtering concept in traditional recommendation systems. STEAM~\cite{STEAM-ARXIV-2026} advances this line of research by introducing structured and evolving agent memory to capture the implicit, diverse, and dynamic preferences reflected in user interactions. By moving beyond a single overwritten preference summary, it enables agents to preserve multi-faceted interests, track preference evolution, and make decisions more consistent with real user choices. Other works like SimUSER~\cite{bougie2025simuser} and SUBER~\cite{corecco2024suber} design cognitive agents with episodic memory, persona grounding, and MDP-based interaction planning, providing realistic behavior logs to support offline evaluation and reinforcement learning policy learning. 
Collectively, these efforts focus on enabling agents to simulate users based on their interaction history and reviews, thereby making decisions that align with real-world user choices.

\textbf{Social Simulation.}
Beyond individual interactions, agent-based frameworks also simulate large-scale social dynamics.
Social simulation aims to leverage user characteristics, contextual information within social networks, and complex mechanisms governing user cognition and decision-making. It employs simulation models capable of reasonably replicating the dynamic properties of historical social behaviors to simulate group-level dynamics, such as the propagation processes of information and emotions\cite{gao2023s3, piao2025agentsociety, jiang2024casevo}.
Regarding explorations of social simulation in recommendation systems, GGBond~\cite{zhong2025ggbond} integrates human-like cognitive agents and dynamic social interactions. It effectively models users' evolving social ties and trust dynamics based on interest similarity, personality compatibility, and structural homogeneity. By responding to recommendations and dynamically updating internal states and social connections, it forms stable multi-round feedback loops. Meanwhile, RecAgent~\cite{RecAgent-TOIS-2025} constructs an interactive sandbox to simulate and study two social scenarios in recommendation systems: information silos and conformity. 
These social simulations contribute rich, multi-agent interaction data that extend beyond isolated user behaviors, providing valuable inputs for building more socially aware and robust recommender systems.

\subsection{Data Unification}
\label{sec3.2:Data Unification}
LLMs provide powerful tools for unifying heterogeneous data across tasks, domains, and modalities, addressing long-standing challenges such as behavior sparsity, domain shift, and representational inconsistency. By leveraging LLMs’ ability to encode diverse inputs into shared semantic spaces and perform prompt-driven generalization, recommender systems can move beyond siloed data processing toward unified architectures. This section highlights four major data-level opportunities: (1) \textbf{multi-domain unification}, enabling cross-domain transfer and cold-start adaptation; (2) \textbf{multi-task unification}, consolidating different recommendation objectives into a single learning framework; (3) \textbf{multi-modal unification}, integrating textual, visual, and behavioral signals into cohesive representations; (4) \textbf{one-model for all}, which builds general-purpose recommender models across scenarios. Together, these approaches reflect the growing trend of using LLMs to unify fragmented data sources and create more holistic, scalable recommendation systems.

\subsubsection{Multi-Domain Data Unification}
\label{sec3.2.1:Multi-Domain Data Unification}
Cross-domain recommendation faces challenges of behavioural sparsity, domain gaps, and representation misalignment. Traditional cross-domain recommendation approaches either model domain overlap from an item perspective or model cross-domain interaction sequences from a user perspective~\cite{MCRPL-TOIS-2024, TCPLP-TOIS-2025}. However, due to the scarcity of overlapping users in practical scenarios, both methods encounter significant challenges. Diffusion-based DMCDR~\cite{DMCDR-KDD-2025} employs a preference encoder to establish preference-guided signals based on source domain interaction histories. These signals are progressively and explicitly injected into user representations to guide the reverse process, enabling cross-domain transfer of user preferences. The powerful representation and reasoning capabilities of LLMs hold promise for bridging semantic relationships between items and deriving users' fine-grained preferences from cross-domain mixed interaction sequences. LLM4CDSR~\cite{LLM4CDSR-SIGIR-2025} employs LLMs to extract semantic representations of items at the semantic level, learning connections between cross-domain items and modelling users' cross-domain interaction sequences. LLMCDSR~\cite{LLMCDSR-TOIS-2025} employs LLMs to comprehend cross-domain information, generating pseudo-interactions for non-overlapping users regarding overlapping items. LLM-RecG~\cite{LLM-RecG-RecSys-2025} tackles zero-shot CDSR by aligning items and sequences without target data, showing promising results. 

In multi-domain recommendation systems, models are prone to being dominated by domains possessing vast datasets. Several approaches have leveraged LLMs to learn domain-general knowledge and inject it into domain-specific recommendation tasks, thereby mitigating this challenge. Uni-CTR~\cite{Uni-CTR-TOIS-2025} employs LLMs to learn hierarchical semantic representations, capturing commonalities across domains. Other approaches train pre-trained language models (PLMs) to combine general textual information with personalized behavioral information from user history sequences, enhancing multi-domain recommendation tasks~\cite{ma2025large, UPSR-TOIS-2025}. CPRec~\cite{ma2025large} mitigates domain distribution gaps by holistically aligning LLMs with general user behavior through a continuous pre-training paradigm. 

In summary, LLM-powered multi-domain unification advances semantic bridging through pretraining, prompt tuning, and modular design, though challenges in scalability and domain interference remain.

\begin{figure*}[!t]
    \centering \includegraphics[width=0.98\textwidth]{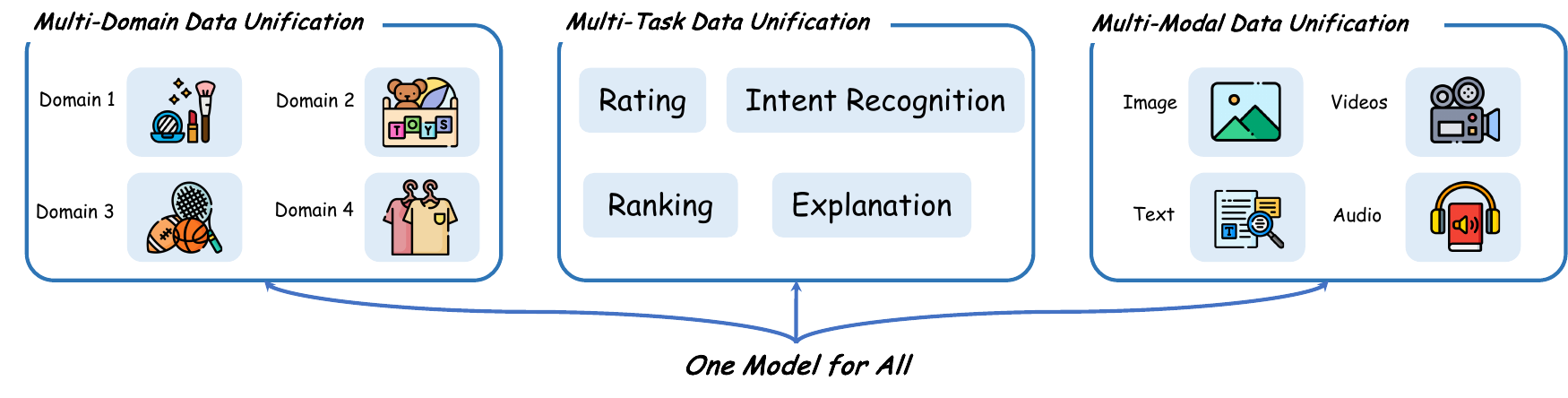}
    \caption{LLM empowered data unification}
    \label{fig:data_stage_2}
\end{figure*}
\subsubsection{Multi-Task Data Unification}
\label{sec3.2.2:Multi-Task Data Unification}
Modern recommender systems face diverse tasks—rating, ranking, explanation, intent recognition—often modeled separately, limiting efficiency and generalization. Multi-task unification integrates these objectives into a single framework for better transfer and scalability.
A common strategy reformulates tasks as language generation. P5 ~\cite{P5-RecSys-2022} pioneered this, unifying recommendation tasks via text-to-text modeling with personalized prompts. GPSD~\cite{GPSD-KDD-2025} further combines generative pretraining with discriminative fine-tuning to improve ranking accuracy. ARTS~\cite{ARTS-TOIS-2025} uses self-prompting for joint prediction and explanation, enhancing interpretability. Efficient Multi-task Prompt Tuning~\cite{MPT-Rec-TOIS-2025} reduces training cost by attaching lightweight prompts to a shared model, enabling scalable deployment.
EcomScriptBench~\cite{ECOMSCRIPT-ACL-2025} offers a multi-task benchmark simulating real-world shopping flows with intent recognition and explanation, supporting realistic evaluation.
In short, multi-task unification boosts performance and flexibility but faces challenges in task interference and parameter efficiency. Future work may focus on modular prompts and unified multimodal training.
\FloatBarrier
\subsubsection{Multi-Modal Data Unification}
\label{sec3.2.3:Multi-Modal Data Unification}
Recommendation involves multiple modalities like text, images, and behavior logs. Traditional methods fuse these separately, but large vision-language models (LVLMs) now enable unified multimodal representations, improving accuracy and interpretability.
UniMP~\cite{UniMP-ICLR-2024} and MQL4GRec~\cite{MQL4GRec-ICLR-2025} unify multimodal inputs into shared semantic spaces, boosting cross-modal generalization. LLaRA~\cite{LLaRA-SIGIR-2024} integrates item IDs and text via hybrid prompting for sequential recommendation, while PAD~\cite{PAD-SIGIR-2025} aligns modalities through a three-stage pretrain-align-disentangle process to enhance cold-start performance.
MSRBench~\cite{MSRBench-WWW-2025} benchmarks LVLM strategies, showing reranking models like GPT‑4o balance performance and cost best, though real-time use remains challenging. 
MLLM-MSR~\cite{harnessing-AAAI-2025} designed an item-summariser based on multimodal large language models (MLLMs) to extract image features of a given item and convert them into text. Supervised fine-tuning (SFT) enables the MLLMs to be employed for multimodal recommendation tasks.
Recent advances explore generative and agentic multimodal recommendation, such as Rec-GPT4V’s zero-shot, multimodal interactions and frameworks for serendipitous recommendations~\cite{Rec-GPT4V-2024}. Some work improves explainability by mapping latent embeddings to readable text. Future systems aim for dynamic, interactive recommendations via multimodal LLMs.
In short, multi-modal unification harnesses foundation models to merge diverse signals into cohesive inputs, advancing personalization and generalization despite challenges in efficiency and deployment.

\subsubsection{One Model for All}
\label{sec3.2.4:One Model for All}
The rapid rise of LLMs is shifting recommender systems from task-specific designs to unified, general-purpose models capable of handling diverse tasks, domains, and modalities, as shown in Fig. \ref{fig:data_stage_2}. This "One Model for All" paradigm leverages foundation models’ capacity for encoding heterogeneous data and performing multi-task inference via prompt-driven frameworks. Early work like P5~\cite{P5-RecSys-2022} reformulates recommendation as a text-to-text generation problem, unifying tasks such as rating prediction and sequential recommendation with personalized prompting. Building on this, M6-Rec~\cite{M6-Rec}removes fixed candidate sets, enabling open-ended multimodal generation combining user behavior, text, and images. UniTRec~\cite{UniTRec-ACL-2023} enhances representation by integrating generative modeling with contrastive learning, improving user intent and item semantic understanding. Similarly, CLLM4Rec~\cite{CLLM4Rec-WWW-2024}incorporates user/item IDs into LLM vocabularies to unify generation and ranking, bridging collaborative filtering with language models.
Beyond recommendations, joint modeling of search and recommendation as sequence generation has shown promising transfer benefits~\cite{penha2024bridging-RecSys-2024}. Multimodal frameworks like vision-language model combine images and text to deliver personalized recommendations and explanations, expanding recommendation toward expressive generation~\cite{rocamonde2023vision}.
From a practical view, A-LLMRec~\cite{A-LLMRec-KDD-2024} integrates pretrained LLMs with collaborative filtering embeddings, using frozen language models and multi-task training to support cold-start, cross-domain, and explainable recommendations in scalable systems.
On representation learning, works like LLM-Rec repurpose pretrained LLMs to generate semantic user and item embeddings, replacing manual features and demonstrating strong cross-domain generalization without fine-tuning.
Beyond unifying tasks and modalities at the data or prompt level, a recent line of work tackles the "One Model for All" challenge from the perspective of \emph{model merging}, i.e., composing a single deployable LLM-based recommender by fusing multiple domain/task specialized LoRA modules in the weight space. RecCocktail~\cite{Hou_Bai_Wu_Liu_Zhang_Liu_Hong_Tang_Wang_2026} trains a reusable "base spirit" LoRA on multi-domain data and merges it with lightweight domain-specific "ingredient" LoRAs, yielding a unified model that inherits both generalization and specialization at no extra inference cost. WeaveRec~\cite{weaverec_WWW_2026} extends model merging to multi-domain fusion in cross-domain sequential recommendation, resolving the conflicts that arise when merging multiple domain-specific LoRAs, and demonstrating that its merging-based approach consistently outperforms naive data-level fusion that simply mixes multi-domain data for joint training. Together, these two works establish model merging as a principled and deployment-friendly route to the "One Model for All" vision.

%% file: model.tex
\section{Model-Level Opportunities}
\label{sec4:Model-Level Opportunities}
\par {Traditional recommender systems primarily use a discriminative framework based on passive user-item interactions for recommendation tasks. However, this approach has notable limitations in both recommendation accuracy and user engagement, as it lacks the ability to incorporate interactive, multi-turn, language-based feedback for refining recommendations.}
\par {The development of generative models offers significant opportunities to address the limitations of traditional recommender systems. LLMs, with their extensive world knowledge, hold substantial potential to enhance recommendation performance. At the same time, the rise of AI-generated content introduces new paradigms, such as the automatic editing of existing items and the generation of new ones. Additionally, diffusion models have gained considerable attention in the recommendation community due to their impressive generative capabilities. This section explores model-level opportunities by categorizing existing generative recommendation systems into three main approaches: (1) \textbf{LLM-Based Generative Recommendation}, (2) \textbf{Large Recommendation Models}, and (3) \textbf{Diffusion-Based Generative Recommendation}.}

\begin{figure*}[!t]
    \centering \includegraphics[width=0.98\textwidth]{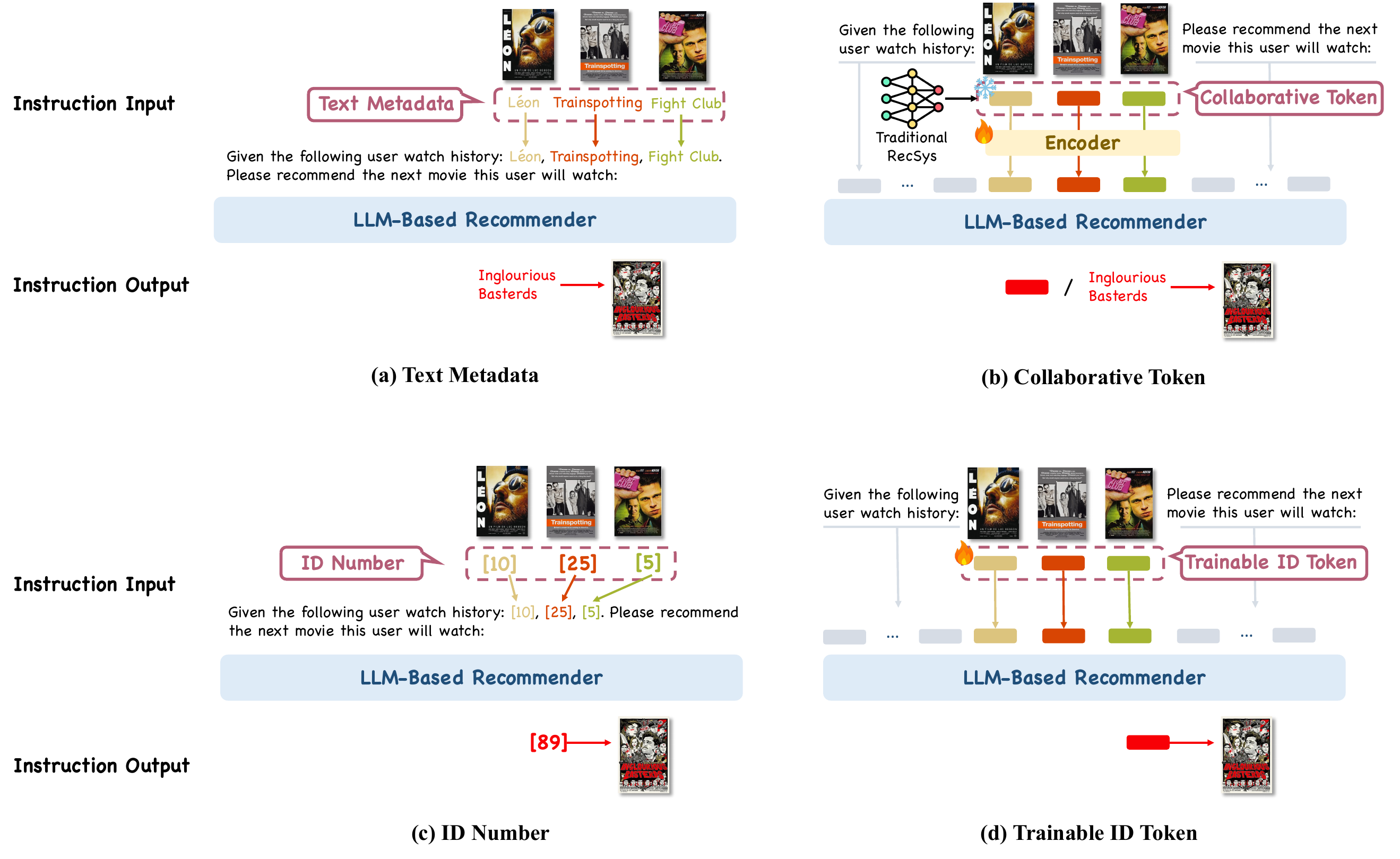}
    \caption{The paradigms of aligning LLMs for recommendation. Inspired by the figure in \cite{Llamarec-2023}.}
    \label{fig:4.2}
\end{figure*}
\subsection{LLM-Based Generative Recommendation}
\label{sec4.1:LLM-Based Generative Recommendation}
With the rapid progress of LLMs, applying them to recommendation has become a major research line. LLM-based generative recommendation leverages pretrained LLMs to produce personalized suggestions via natural-language prompts or lightweight fine-tuning, tapping the broad world knowledge encoded in the models. Existing studies can be grouped into three strands: (i) Pretrained LLMs for recommendation, which explore prompt design and tool use to solve recommendation tasks without heavy retraining; (ii) Aligning LLMs to recommendation, which tackles the mismatch between generic language modeling objectives and the goals of ranking, personalization, and domain constraints; and (iii) Training objectives and inference, which develop fine-tuning losses, preference-optimization methods, and efficient decoding to improve accuracy and efficiency.

\subsubsection{Pretrained LLMs Recommendation}
\label{sec4.1.1:Pretrained LLMs Recommendation}
\par The utilization of pretrained LLMs for recommendation relies on prompt design and in-context learning, so the models can be applied without extra fine-tuning. A key benefit is that LLMs bring broad world knowledge and strong semantic understanding, which helps in cold-start and cross-domain settings. Research in this line follows two complementary routes: (i) LLM-as-Enhancer~\cite{sileo2022zero-ECIR-2022, hou2024large-ECIR-2024, kang2023llms-2023, liu2023chatgpt-CIKM-2023, harrison2023zero-ICDMW-2023}. LLMs are used to rewrite user/item profiles and interaction histories into natural-language features, which are then fed to or combined with collaborative filtering, sequential models, or re-rankers. This improves explainability, user interaction, and sometimes long-tail coverage. (ii) LLM-as-Recommender~\cite{Chat-REC-2023}. With task-specific prompts or templates, a pretrained LLM directly generates recommendations (e.g., item titles or IDs) and can operate in zero-shot mode across scenarios. This idea extends to multimodal settings~\cite{wang2025hanorec, wang2026mllmrec-r1}, where MLLMs take both text and images (e.g., posters) as input to produce zero-shot recommendations. These approaches show how pretrained LLMs can either enhance existing pipelines or serve as generators, offering flexible ways to deploy LLM capability in recommender systems~\cite{harrison2023zero-ICDMW-2023}.

\subsubsection{Aligning LLMs for Recommendation.} 
\label{sec4.1.2:Aligning LLMs for Recommendation}
\par Aligning LLMs for recommendation adapts a pretrained model to recommender objectives and evaluation. While LLMs offer strong semantics and zero-/few-shot use, their pretraining ignores click/engagement signals, top-k ranking, long-tail coverage, and exposure bias, so direct use often trails specialized models. To bridge this gap, fine-tuning on recommendation-specific data has emerged as a promising approach to align the training signal with language models. During fine-tuning, these works generate user/item profiles to present the LLM with a structured, compact, and consistent collaborative signal. By how the user/item profile injects structure, approaches include: text prompting, collaborative signal, and item tokenization. In Table \ref{tab:text-prompt}, \ref{tab:collaborative-signal}, \ref{tab:item-tokenization}, we summarize representative works of these methods, respectively.

\par \textbf{Text prompting based methods.} 
Text prompting based methods build the user profile entirely in natural language by combining the task description with the user’s chronological interaction history. The framework is shown in Fig.~\ref{fig:4.2} (a). 
Early work~\cite{CALRec-RecSys-2024, BiLLP-SIGIR-2024, PO4ISR-SIGIR-2024} mainly feeds the sequence of consumed items, but such inputs may miss fine-grained preference signals.
To enrich the profile, later studies explicitly model preferences within the history: TALLRec~\cite{TALLRec-RecSys-2023} inserts explicit preference statements into prompt templates and uses LoRA for lightweight adaptation; LlamaRec~\cite{Llamarec-2023} first contracts the candidate set with a sequential recommender to provide a focused context; LRD~\cite{LRD-SIGIR-2024} couples LLM knowledge with a variational autoencoder to reveal latent item relations. Some works also adopt preference optimization to keep profiles flexible—LettinGo~\cite{LettinGo-KDD-2025} uses direct preference optimization to adapt the model to user preferences without rigid supervised targets. 
Beyond static histories, recent studies incorporate dynamic feedback to track evolving interests~\cite{RecGPT-ACL-2024, EXP3RT-SIGIR-2025, Reason4Rec-Arxiv-2025}; for instance, Reason4Rec~\cite{Reason4Rec-Arxiv-2025} leverages user reviews to extract preferences and salient item attributes, helping the model judge the match between user interests and candidates.
While text prompting makes recommendations possible with textual descriptions alone, it lacks explicit collaborative signals. As a result, performance can fall short in settings where inter-item dependencies and collaborative patterns are critical for ranking quality.

\par \textbf{Collaborative signal based methods.} 
To capture inter-item relations that plain text misses, these methods inject collaborative signals into the user/item profile so the LLM sees both semantics and relational knowledge (Fig.~\ref {fig:4.2} (b)). 
Existing studies can be broadly categorized into three directions: (i) LLM-augmented representation for CF models, which map Collaborative signal and semantic representations into a shared space and fuse them to form enriched profiles. Methods such as~\cite{iLoRA-NeurIPS-2024, LLM-ESR-NeurIPS-2024, LLaRA-SIGIR-2024, PAD-SIGIR-2025, BinLLM-ACL-2024, CoLLM-TKDE-2025, E4SRec-WWW-2024, SeRALM-SIGIR-2024, A-LLMRec-KDD-2024, RLMRec-WWW-2024, RecLM-ACL-2025} concatenate the two sources of information to construct enriched user profiles, leveraging their complementary strengths. (ii) LLM-assisted summarization for CF models, which distills user preferences into compact textual summaries that serve as auxiliary inputs to traditional recommenders. For example, CORONA\cite{CORONA-SIGIR-2025} combines LLM reasoning with a GNN in a coarse-to-fine pipeline, and HyperLLM~\cite{HyperLLM-SIGIR-2025} uses LLM-generated summaries to enhance collaborative models. CoRAL~\cite{CoRAL-KDD-2024} reformulates collaborative signals as explicit sentences (e.g., “User A also prefers X, Y, Z”), making them more interpretable and directly usable by LLMs.
While collaborative-signal methods enrich user profiles with interaction-derived information, dense embeddings are not natively interpretable for LLMs. In practice, they often require projection or verbalization to bridge representation spaces, and this mapping may introduce a potential gap between collaborative and textual semantics.

\begin{table*}[t]

\centering
\caption{Summary of the representative text prompting-based recommendation methods.}
\resizebox{\textwidth}{!}{
\begin{tabular}{l|cccc|l}
\hline
\multicolumn{1}{c|}{\multirow{2}{*}{Methods}} & \multicolumn{4}{c|}{User formulation} & \multicolumn{1}{c}{\multirow{2}{*}{Backbone}} \\ \cline{2-5}
\multicolumn{1}{c|}{} & \multicolumn{1}{c}{Task description} & \multicolumn{1}{c}{Historical interactions} & \multicolumn{1}{c}{Profile} & \multicolumn{1}{c|}{Feedback} & \multicolumn{1}{c}{} \\ \hline
Chat-Rec~\cite{Chat-REC-2023}            & ranking & history interactions & \checkmark & -- & GPT-3.5 \\
TALLRec~\cite{TALLRec-RecSys-2023}       & preference classification & user preference & -- & -- & LLaMA-7B \\
LlamaRec~\cite{Llamarec-2023}            & retrieval, ranking & history interactions & -- & -- & LLaMA2-7B \\
LRD~\cite{LRD-SIGIR-2024}                & ranking & history interactions         & -- & -- & GPT-3.5 \\
ReLLa~\cite{ReLLa-WWW-2024}              & ranking & history interactions & \checkmark & -- & Vicuna-7B \\
CALRec~\cite{CALRec-RecSys-2024}         & ranking & history interactions & -- & -- & PaLM-2 XXS \\
BiLLP~\cite{BiLLP-SIGIR-2024}            & long-term Interactive  & history interactions, reward model & -- & -- & GPT-3.5, GPT-4, LLaMA2-7B \\
PO4ISR~\cite{PO4ISR-SIGIR-2024}          & Ranking & history interactions & -- & -- & LLaMA2-7B \\
LLM-TRSR~\cite{LLM-TRSR-WWW-2024}        & Ranking & history interactions & \checkmark & -- & LLaMA2-7B \\
RecGPT~\cite{RecGPT-ACL-2024}            & Ranking & history interactions, user preference & -- & \checkmark & RecGPT-7B \\
KAR~\cite{KAR-RecSys-2024}               & Ranking & history interactions, user preference & \checkmark & -- & GPT-3.5 \\
LLM4CDSR~\cite{LLM4CDSR-SIGIR-2025}      & Ranking  & history interactions & \checkmark & \checkmark & GPT-3.5, GLM4-Flash \\
EXP3RT~\cite{EXP3RT-SIGIR-2025}          & rating prediction & history interactions & \checkmark & \checkmark & LLaMA3-8B \\
SERAL~\cite{SERAL-KDD-2025}              & retrieval, ranking  & history interactions & \checkmark & -- & Qwen2-0.5B \\
LettinGo~\cite{LettinGo-KDD-2025}        & Ranking & history interactions & \checkmark & -- & LLaMA3-8B \\
Reason4Rec~\cite{Reason4Rec-Arxiv-2025}        & Rating Prediction & history interactions, user preference & -- & \checkmark & LLaMA3-8B \\
InstructRec~\cite{InstructRec-TOIS-2025} & Ranking & history interactions & \checkmark & -- & Flan-T5-XL \\
Uni-CTR~\cite{Uni-CTR-TOIS-2025}         & Rating Prediction & history interactions, user preference & \checkmark & -- & DeBERTaV3-large \\
BIGRec~\cite{BIGRec-TORS-2025}           & Ranking & history interactions & -- & -- & LLaMA-7B \\
UPSR~\cite{UPSR-TOIS-2025}               & Ranking & history interactions & -- & -- & T5, FLAN-T5 \\ \hline
\end{tabular}
}
\label{tab:text-prompt}
\end{table*}

\begin{table*}[ht]

\centering
\caption{Summary of the representative collaborative signal-based methods.}
\resizebox{\textwidth}{!}{
\begin{tabular}{l|cccc|l|l}
\hline
\multirow{2}{*}{\textbf{Methods}} & \multicolumn{4}{c|}{\textbf{User Formulation}} & \multicolumn{1}{c|}{\multirow{2}{*}{\textbf{Combining Method}}} & \multicolumn{1}{c}{\multirow{2}{*}{\textbf{Backbone}}} \\ \cline{2-5}
                                  & \textbf{Task Description} & \textbf{Historical Interactions} & \textbf{Profile} & \textbf{Feedback} & \multicolumn{1}{c|}{} & \multicolumn{1}{c}{} \\ \hline
iLoRA~\cite{iLoRA-NeurIPS-2024}    & Ranking              & history interactions                     & -                  & -                  & Concatenation & GPT-3.5  \\ 
LLM-ESR~\cite{LLM-ESR-NeurIPS-2024} &  Ranking              & history interactions                     & -                  & -                  & Concatenation & LLaMA2-7B \\ 
LLaRA~\cite{LLaRA-SIGIR-2024}       & Ranking              & history interactions                     & -                  & -                  & Concatenation & LLaMA2-7B \\ 
A-LLMRec~\cite{A-LLMRec-KDD-2024}   & Ranking              & history interactions                     & -                  & -                  & Concatenation & OPT-6.7B \\ 
RLMRec~\cite{RLMRec-WWW-2024}       & Ranking              & history interactions, user preference                     & -                  & \checkmark         & Concatenation & GPT-3.5  \\ 
CoRAL~\cite{CoRAL-KDD-2024}         & Ranking              & history interactions, user preference                     & -                  & \checkmark         & Retrieval-Augmented & GPT-4 \\ 
BinLLM~\cite{BinLLM-ACL-2024}       & Ranking              & history interactions, user preference                     & -                  & \checkmark         & Concatenation & Vicuna-7B \\ 
E4SRec~\cite{E4SRec-WWW-2024}       & Ranking              & history interactions                     & -                  & -                  & Concatenation & Vicuna-7B \\ 
SeRALM~\cite{SeRALM-SIGIR-2024}     & Ranking              & history interactions                      & -                  & -                  & Concatenation & LLaMA2-7b \\ 
CORONA~\cite{CORONA-SIGIR-2025}     & Ranking              & history interactions                     & \checkmark         & -                  & Pipeline Integration & GPT-4o-mini \\ 
HyperLLM~\cite{HyperLLM-SIGIR-2025} & Ranking              & history interactions                     & -                  & -                  & Pipeline Integration & LLaMA3-8B \\ 
RecLM~\cite{RecLM-ACL-2025}         & Ranking              &  history interactions                     & \checkmark         & -                  & Concatenation & LLaMA2-7b \\ 
CoLLM~\cite{CoLLM-TKDE-2025}        & Ranking              & history interactions                     & -                  & -                  & Concatenation & Vicuna-7B \\ 
PAD~\cite{PAD-SIGIR-2025}           & Ranking                       & history interactions, user preference                     & -                  & -                  & Concatenation & LLaMA3-8B \\ 
IDP~\cite{IDP-TOIS-2025}            & Ranking              & history interactions, user preference                     & -                  & -                  & Concatenation & T5 \\ \hline
\end{tabular}
}
\label{tab:collaborative-signal}
\end{table*}


\begin{table*}

\centering
    \caption{Summary of the representative item tokenization-based methods.}
\resizebox{\textwidth}{!}{
\begin{tabular}{l|ccc|l}
\hline
\multicolumn{1}{c|}{\multirow{2}{*}{Methods}} & \multicolumn{3}{c|}{User formulation}                                  & \multicolumn{1}{c}{\multirow{2}{*}{Backbone}} \\ \cline{2-4}
\multicolumn{1}{c|}{}                         & Task description & Historical interactions                  & Token types                           & \multicolumn{1}{c}{}                         \\ \hline
P5~\cite{P5-RecSys-2022}                      & Ranking         & historical interactions, user preference  & ID-based tokenization                 & Transformer                                  \\
CLLM4Rec~\cite{CLLM4Rec-WWW-2024}             & Ranking         & historical interactions                   & ID-based tokenization                 & GPT-2                                        \\
BIGRec~\cite{BIGRec-TORS-2025}                & Ranking         & historical interactions                   & Text-based tokenization               & LLaMA-7B                                     \\
M6~\cite{M6-Rec}                              & Retrieval, Ranking & historical interactions                & Text-based tokenization               & M6                                           \\
IDGenRec~\cite{IDGenRec-SIGIR-2024}           & Ranking         & historical interactions                   & Text-based tokenization               & BERT4Rec                                     \\
TIGER~\cite{TIGER-NeurIPS-2023}               & Ranking         & historical interactions                   & Codebook-based tokenization           & T5                                           \\
RPG~\cite{RPG-KDD-2025}                       & Ranking         & historical interactions                   & Codebook-based tokenization           & LLaMA-2-7B                                   \\
LC-Rec~\cite{LC-Rec-ICDE-2024}                & Ranking         & historical interactions                   & Codebook-based tokenization           & LLaMA-2-7B                                   \\
ActionPiece~\cite{ActionPiece-ICML-2025}      & Retrieval       & historical interactions                   & Codebook-based tokenization           & LLaMA-2-7B                                   \\
LETTER~\cite{LETTER-CIKM-2024}                & Ranking         & historical interactions                   & Codebooks with collaborative signals  & LLaMA-7B                                     \\
TokenRec~\cite{tokenrec-TKDE-2025}            & Retrieval       & historical interactions                   & Codebooks with collaborative signals  & T5-small                                     \\
SETRec~\cite{SETRec-SIGIR-2025}               & Ranking         & historical interactions                   & Codebooks with collaborative signals  & T5, Qwen                                     \\
CCFRec~\cite{CCFRec-KDD-2025}                 & Ranking         & historical interactions                   & Codebooks with collaborative signals  & LLaMA-2-7B                                   \\
LLM2Rec~\cite{LLM2Rec-KDD-2025}               & Ranking         & historical interactions                   & Codebooks with collaborative signals  & LLaMA-2-7B                                   \\
SIIT~\cite{SIIT-arxiv-2024}                   & Retrieval       & historical interactions                   & Self-adaptive tokenization            & LLaMA-2-7B                                   \\ \hline
\end{tabular}
}
\label{tab:item-tokenization}
\end{table*}

\begin{table*}[htbp]
\caption{Unified training objectives for LLM-based generative recommendation.}
\centering
\begin{tabular}{llc}
\toprule
\textbf{Category} & \textbf{Representative Works} & \textbf{Formula} \\ 
\midrule

\multirow{2}{*}{\textbf{Supervised Fine-Tuning}} 
& P5 (RecSys'22) LGIR (AAAI'24) & \multirow{2}{*}{$-\log \pi_{\theta}(y^{+} \mid x)$} \\
& LLM-Rec (TOIS'25) RecRanker (TOIS'25) & \\[4pt]

\multirow{2}{*}{\textbf{Self-Supervised Learning}}  
& FELLAS (TOIS'24) & \multirow{2}{*}{$-\log \frac{\exp\!\big(\mathrm{sim}(y^{+}, y^{-})/\tau\big)}{\sum_{y \in \mathcal{N}} \exp\!\big(\mathrm{sim}(y^{+}, y)/\tau\big)}$} \\
& HFAR (TOIS'25) & \\[4pt]

\multirow{2}{*}{\textbf{Reinforcement Learning}} 
& LEA (SIGIR'24) & \multirow{2}{*}{$- \Big[ r_{\phi}(x, y^{+}) - \beta\, D_{\mathrm{KL}}\!\big(\pi_{\theta}(y \mid x)\,\|\,\pi_{\mathrm{ref}}(y \mid x)\big) \Big]$} \\
& RPP (TOIS'25) & \\[4pt]

\multirow{2}{*}{\textbf{Direct Preference Optimization}} 
& LettinGo (KDD'25) RosePO (ArXiv'24) & \multirow{2}{*}{$-\log \sigma\!\left(\beta \log \frac{\pi_{\theta}(y^{+}\!\mid x)}{\pi_{\mathrm{ref}}(y^{+}\!\mid x)}-\log \frac{\pi_{\theta}(y^{-}\!\mid x)}{\pi_{\mathrm{ref}}(y^{-}\!\mid x)}\right)$} \\ 
& SPRec (WWW'25) & \\[6pt]

\multicolumn{3}{l}{
\parbox{0.95\textwidth}{\footnotesize
\textit{Notation:} $x$ is user profile/context; $y^+$/$y^-$ is preferred/rejected item; 
$\pi_{\theta}$ is policy model; $\pi_{\mathrm{ref}}$ is reference model; 
$\mathrm{sim}(\cdot,\cdot)$ is similarity; $\tau$ is temperature; $\mathcal{N}$ is negative set; 
$r_{\phi}$ is reward; $\beta$ is penalty/scale; $D_{\mathrm{KL}}$ is KL divergence; 
$\sigma$ is sigmoid.
}}
\\
\bottomrule
\end{tabular}

\label{tab:trainobjective}
\end{table*}

\par \textbf{Item tokenization based methods.} 
To narrow the gap between collaborative and textual semantics, this line of work maps items into the LLM’s vocabulary by assigning identifiable tokens that the model can generate autoregressively (Fig.~\ref{fig:4.2} (c) and (d)). The key challenge is to design identifiers that are both efficient and semantically meaningful, balancing textual semantics with collaborative knowledge. Existing work can be grouped into five directions: 
(i) ID-based tokenization: The simplest approach assigns a special token to each user or item~\cite{P5-RecSys-2022, CLLM4Rec-WWW-2024}. While conceptually straightforward, this strategy is infeasible at scale due to vocabulary explosion and its inability to encode semantics, making cold-start generalization difficult. 
(ii) Text-based tokenization: To introduce semantics, some methods~\cite{BIGRec-TORS-2025, M6-Rec, IDGenRec-SIGIR-2024, dmrec-SIGIR-2025, proex-KDD-2026, zhang2026promax} use textual attributes such as titles and descriptions. However, these are often too lengthy for autoregressive generation and fail to capture collaborative knowledge, limiting effectiveness in practice. 
(iii) Codebook-based tokenization: A more compact alternative represents items as sequences of discrete tokens from a shared vocabulary~\cite{TIGER-NeurIPS-2023, RPG-KDD-2025, LC-Rec-ICDE-2024}. This reduces vocabulary size and avoids overly long sequences while retaining partial semantics. For instance, Semantic IDs cluster items with similar attributes to share tokens, and ActionPiece~\cite{ActionPiece-ICML-2025} compresses feature patterns into tags. Yet, these methods struggle to balance textual and collaborative semantics. 
(iv) Codebooks with collaborative signals: To bridge this gap, recent studies integrate CF signals directly into tokenization. LETTER\cite{LETTER-CIKM-2024} combines RQ-VAE with contrastive alignment to regulate semantics and collaboration; TokenRec\cite{tokenrec-TKDE-2025} quantizes masked user/item embeddings into discrete tokens, smoothly incorporating high-order collaborative knowledge; SETRec\cite{SETRec-SIGIR-2025} encodes histories as unordered sets; CCFRec\cite{CCFRec-KDD-2025} enhances modeling with code masking and alignment; and LLM2Rec~\cite{LLM2Rec-KDD-2025} injects CF signals via supervised fine-tuning and embedding modeling. 
(v) Self-adaptive tokenization via LLMs: The latest direction allows LLMs themselves to refine identifiers during training. SIIT~\cite{SIIT-arxiv-2024}, for example, enables self-tuning of item tokens, aligning them with internal representations and mitigating inconsistencies caused by external tokenizers.
Item tokenization provides a scalable solution to unify textual and collaborative semantics, but designing tokens that balance both efficiently remains an open challenge.

\subsubsection{Training Objective \& Inference}
\label{sec4.1.3:Training Objective and Inference}
\par While aligning LLMs with recommendation data improves their adaptability, model performance still hinges on how training and inference are formulated. To this end, researchers have proposed a series of objectives and strategies specifically designed for recommendation scenarios.

\par \textbf{Training objective.} 
In recommender systems, the training objective is typically next-item prediction, where models infer the most probable subsequent item given a user’s interaction history. To align large language models (LLMs) with this objective, prior work explores four training paradigms, each addressing a distinct limitation. In Table \ref{tab:trainobjective}, we list representative works and formulas for these four training paradigms.
\begin{itemize}[leftmargin=*]
    \item Supervised Fine-Tuning (SFT). To learn next-item prediction, LLMs are fine-tuned with predefined templates that specify the task, encode interaction histories, and standardize outputs~\cite{P5-RecSys-2022, LLM-Rec-TOIS-2025, RecRanker-TOIS-2025}. For example, P5 \cite{P5-RecSys-2022} designs prompts for five representative tasks, and LGIR \cite{LGIR-AAAI-2024} augments SFT with a GAN-based module to improve robustness in few-shot settings. Despite strong results, SFT mainly learns from positive pairs and lacks explicit negatives, making it harder to learn ranking margins and to adapt to evolving user preferences.
    \item Self-Supervised Learning (SSL). SSL-based methods~\cite{FELLAS-TOIS-2024, HFAR-TOIS-2025, EasyRec-arxiv-2024} reduces reliance on manual templates by generating auxiliary training signals. For example, EasyRec~\cite{EasyRec-arxiv-2024} builds a text–behavior alignment objective that combines contrastive learning with collaborative language-model tuning, enabling the LLM to separate candidates without explicit labels and improving zero-shot generalization.
    \item Reinforcement learning (RL). RL-based methods introduces reward-driven optimization over ranked sessions to model negatives and handle non-differentiable metrics. LEA~\cite{LEA-SIGIR-2024} learns user states with task-specific rewards, and RPP~\cite{RPP-TOIS-2025} shapes rewards from downstream metrics to directly optimize recommendation goals. While effective, RL often needs large-scale feedback and can be costly and unstable to train.
    \item Preference Optimization (PO). To avoid training a reward model and reduce the instability of RL, PO-base~\cite{LettinGo-KDD-2025, xu2026multi,xu2026vc, Rosepo-arxiv-2024, sprec-WWW-2025} methods directly optimizes on preference pairs. In recommendation, RosePO~\cite{Rosepo-arxiv-2024} tailors preference construction and applies a DPO-style objective to align the model with ranking goals without explicit rewards. SPRec~\cite{sprec-WWW-2025} adopts a self-play pipeline to stabilize training and strengthen preference alignment without a separate reward model.
\end{itemize}

\par \textbf{Inference}. In recommender systems, direct inference lets the LLM directly generate the Top-K items or score candidates one by one, yielding the simplest pipeline and requiring no re-ranker or reward model. However, it is prompt-sensitive, hard to control for stability and diversity, lacks explicit ranking signals, and long histories inflate prompts and memory overhead. These limitations motivate two complementary lines of work: inference reranking and inference acceleration.

\begin{itemize}[leftmargin=*]
    \item Reranking. Improve output quality by injecting stronger ranking signals at inference: RecRanker~\cite{RecRanker-TOIS-2025} adopts a two-stage pipeline (candidate retrieval followed by LLM-based reranking) with a hybrid objective that integrates pairwise and applies position shifting to mitigate input input bias: LLM4Rerank~\cite{LLM4Rerank-WWW-2025} frames inference as multi-node, multi-hop reasoning with goal guidance and history pools to trade off accuracy, diversity, and fairness; GFN4Rec~\cite{GFN4Rec-KDD-2023} employs GFlowNets to autoregressively generate lists, improving coverage and diversity.
    \item Acceleration. Reduce latency and memory by cutting LLM usage, shortening inputs, and speeding decoding: FELLAS limits the LLM to produce item/sequence embeddings while a lightweight model makes final predictions~\cite{FELLAS-TOIS-2024}; Prompt Distillation (GenRec~\cite{GenRec-RecSys-2023}) compresses long histories and distills to a small model so the LLM is triggered only when needed; AtSpeed~\cite{AtSpeed-ICLR-2025} applies speculative decoding and tree-based attention to achieve $2$–$2.5\times$ speedups while preserving Top-K consistency.
\end{itemize}

\FloatBarrier

\begin{figure*}[!t]
    \centering \includegraphics[width=0.98\textwidth]{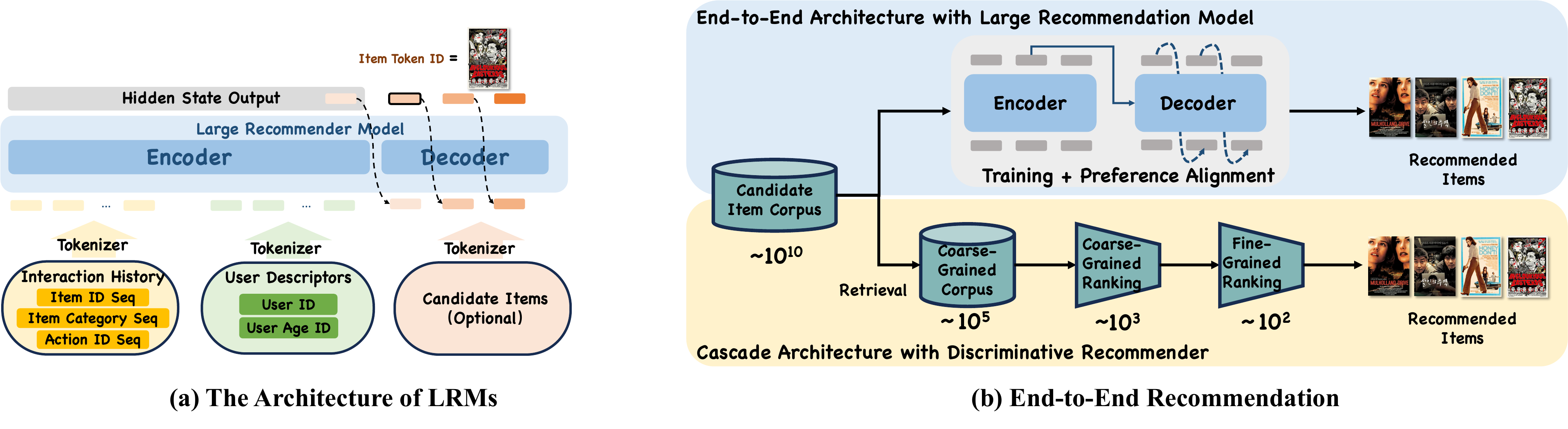}
    \caption{Illustration of two research directions of large recommendation models. (a) The Architecture of LRMs, and (b) End-to-End Recommendation.} 
    \label{fig:LRM}
\end{figure*}

\subsection{Large Recommendation Model}
\label{sec4.2:Large Recommendation Model}
The success of LLMs has established compelling scaling laws: model performance improves predictably with increased compute, data, and parameters following power-law relationships. LLM-based recommenders discussed above inherit the benefits of pre-trained language models whose scaling laws have been well-established, but this comes with the requirement of reformulating user behavior data as natural language sequences. Recently, a parallel research direction has emerged that forgoes LLM adaptation in favor of establishing native scaling laws tailored specifically for recommendation tasks~\cite{HSTU-ICML-2024,OneRec-2025,guo-OneSug-2025,TIGER-NeurIPS-2023,GenRec-arxiv-2025,OneRec-v2-arxiv-2025}. Rather than borrowing from language models, these approaches design specialized architectures optimized directly for user behavior data. We refer to this emerging paradigm as Large Recommendation Models (LRMs).

LRMs have emerged as an active research area in industry in 
recent years, enabled by the massive scale of user behavior data available in 
industrial settings. In recent few years, industrial recommendation teams have faces two critical bottlenecks: First, diminishing returns from discriminative recommendation models' complexity. Increasingly sophisticated architectures no longer yield significant performance improvements.  For instance, while Alibaba's Deep Interest Network (DIN)\cite{DIN-KDD-2018} achieved substantial gains, subsequent models like MIMN~\cite{pi2019practice} showed progressively marginal improvements—a pattern observed across all major industry teams.
Second, escalating costs of cascaded architectures. Multi-stage recommendation pipelines in industry, typically consisting of retrieval, coarse ranking, fine ranking, and re-ranking stages, have become increasingly complex, resulting in higher maintenance overhead and growing communication and caching costs between stages.
To address these bottlenecks, recent industrial research has increasingly turned to LRMs along two primary directions as shown in Fig. \ref{fig:LRM}. First, to overcome the diminishing returns from model complexity, researchers are developing novel generative training paradigms that fundamentally rethink how recommendation models learn from data. Second, to mitigate the costs of cascaded architectures, efforts focus on leveraging LRMs to construct end-to-end recommendation frameworks that bypass the traditional multi-stage pipeline altogether.
In Table \ref{tab:lrm}, we summarize some representative large recommendation models.

\FloatBarrier
\begin{table*}
\centering
    \caption{Summary of the representative large recommendation model}
    \label{tab:lrm}
\resizebox{\textwidth}{!}{
\begin{tabular}{l|cc|c|l}
\hline
\multicolumn{1}{c|}{\multirow{2}{*}{Methods}}              & \multicolumn{2}{c|}{User Formulation}                                   & \multirow{2}{*}{Architectures} & \multicolumn{1}{c}{\multirow{2}{*}{Backbone}} \\ \cline{2-3}
\multicolumn{1}{c|}{}                                      & Task                             & Historical Interactions               &                                & \multicolumn{1}{c}{}                          \\ \hline
LEARN~\cite{LEARN-AAAI-2025}         & Ranking                          & history interactions, user preference & Cascaded                       & Baichuan2-7B, Transformer                     \\
HLLM~\cite{chen-HLLM-2024}           & Retrieval, Ranking               & history interactions                  & Cascaded                       & TinyLlama-1.1B, Baichuan2-7B                  \\
KuaiFormer~\cite{liu2024-kuaiformer} & Retrieval                        & history interactions                  & Cascaded                       & Stacked Transformer                           \\
SRP4CTR~\cite{SRP4CTR-CIKM-2024}     & Ranking                          & history interactions, user preference & Cascaded                       & FG-BERT                                       \\
HSTU~\cite{HSTU-ICML-2024}           & Ranking                          & history interactions                  & Cascaded                       & Transformer                                   \\
MTGR~\cite{MTGR-arxiv-2025}        & Ranking                          & history interactions                  & Cascaded                       & Transformer                                   \\
UniROM~\cite{Qiuone-2025}            & Ranking                          & history interactions                  & End-to-End                     & RecFormer                                     \\
URM~\cite{URM-2025}                  & Ranking                          & history interactions, user preference & End-to-End                     & BERT                                          \\
OneRec~\cite{OneRec-2025}            & Generative Retrieval and Ranking & history interactions, user preference & End-to-End                     & Transformer                                   \\
OneSug~\cite{guo-OneSug-2025}        & Generative Retrieval and Ranking & history interactions, user preference & End-to-End                     & Transformer                                   \\
EGA-V2~\cite{zheng-EGAV2-2025}       & Generative Retrieval and Ranking & history interactions, user preference & End-to-End                     & Transformer                                   \\ \hline
\end{tabular}
}
\end{table*}

\par \textbf{The Scaling Law of LRMs.} 
Meta's proposed HSTU~\cite{HSTU-ICML-2024} is groundbreaking work in large recommendation models, validating that the scaling law of LLM is equally applicable to recommendation systems. HSTU transforms the traditional discriminative CTR prediction task into a generative sequence modeling task. For each user, it unified multiple pointwise user-item interaction samples into a single user behavior sequence containing interacted items, interaction behaviors, user and items' categorical features. HSTU adopts a causal autoregressive modeling approach that takes ultra-long user sequences as input, unifying the retrieval and ranking tasks into a sequence generation problem. The output is a probability distribution over candidate items, supporting multi-task joint training. HSTU employs sequence lengths ranging from 1024 to 8192, which far exceeds the length that traditional discriminative models can handle. Moreover, from a performance perspective, there is a trend showing that longer sequences and more complex models lead to better results. HSTU validates that as model scale increases, its performance continues to improve, whereas traditional discriminative recommendation models plateau. Ultimately, HSTU reaches a parameter scale of 1.5 trillion, while discriminative recommendation models stagnate in effectiveness at around 200 billion parameters. Based on HSTU, several major tech companies have successively proposed LRMs. For example, Meituan proposed MTGR~\cite{MTGR-arxiv-2025}, a generative ranking framework. MTGR incorporates cross features commonly used in discriminative recommendation systems to prevent information loss in the generative architecture. For sequence encoding, it combines Group LayerNorm and a dynamic hybrid masking strategy to enhance HSTU's learning effectiveness.
Redbook proposed GenRank~\cite{GenRec-arxiv-2025}, applied to their fine-grained ranking applications. HSTU concatenates user-interacted items and their corresponding actions sequentially in the input sequence, which doubles the sequence length. GenRank treats items as positional information and focuses on iteratively predicting actions associated with items, significantly reducing the input sequence length, making it suitable for resource-sensitive ranking scenarios.

\par \textbf{End-to-End Recommendations.}
The new paradigm of large recommendation models also makes it possible to unify the recommendation framework. Simplifying the recommendation framework can reduce engineering costs, and merging cascade structures can unify model optimization objectives, bringing performance gains. Kuaishou's large recommendation model OneRec~\cite{OneRec-2025,OneRec-v2-arxiv-2025} has made an excellent attempt in this direction. OneRec uses an end-to-end generative recommendation model to replace the traditional retrieval-coarse ranking-fine ranking cascade architecture. The entire OneRec solution achieved a significant 1.68\% improvement in the key online metric of total watch time. The system's computational resource utilization increased from 11\% to 28.8\%. Since there's no need for inter-layer communication and data caching, OneRec's runtime cost is only 10.6\% of the cascade architecture, demonstrating that generative models have found new paradigms for both performance improvement and engineering optimization in recommendation systems. In terms of model architecture, OneRec adopts an encoder-decoder structure and uses the MoE architecture to expand model capacity and enhance user interest modeling capabilities. For generation methods, unlike traditional pointwise prediction, OneRec proposes a session-based generation approach that generates entire recommendation lists to better capture contextual information. In terms of training, OneRec adds a preference alignment phase, using Direct Preference Optimization (DPO) with a reward model to generate preference data and optimize generation results.
OneSug~\cite{guo-OneSug-2025} extends this idea to query recommendation, unifying prefix, context, and history with RQ-VAE–based semantic IDs and Reward-Weighted Ranking to align with graded feedback. EGA-V2~\cite{zheng-EGAV2-2025} pushes this direction further by introducing hierarchical tokenization and multi-token prediction, integrating user interest modeling, POI generation, creative selection, ad allocation, and payment computation into a single generative framework. Together, these advances illustrate how end-to-end models are evolving from unifying retrieval and ranking to fully generative paradigms that integrate diverse recommendation tasks.

\begin{figure*}[!t]
    \centering \includegraphics[width=0.98\textwidth]{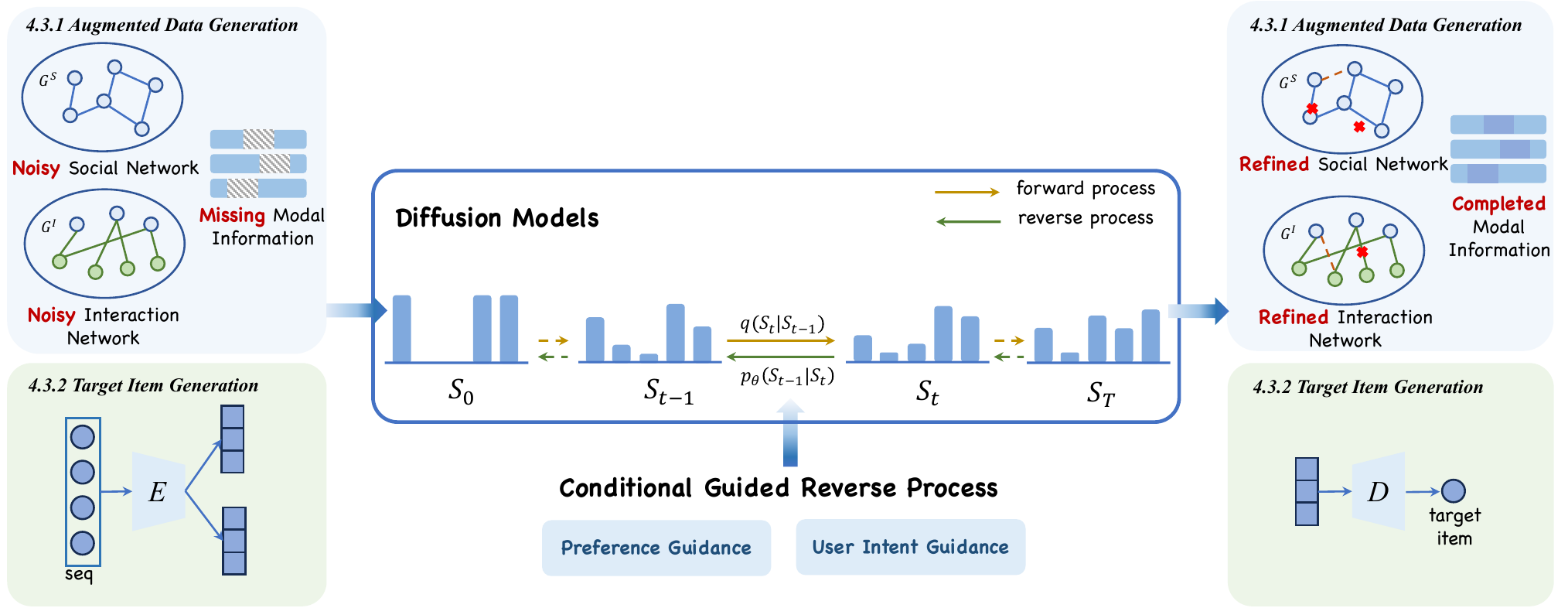}
    \caption{Illustration of two types of diffusion-based generative recommendation. (a) Augmented Data Generation, and (b) Target Item Generation.}
    \label{fig:4.3}
\end{figure*}

\subsection{Diffusion-Based Generative Recommendation}
\label{sec4.3:Diffusion-Based Generative Recommendation}
With the remarkable success of diffusion models (DM) in image synthesis, recent research has explored extending diffusion models to various recommendation tasks. DM-based methods can be categorized into two main types: approaches for augmented data generation and approaches for target item generation. 
\subsubsection{Augmented Data Generation} 
The denoising characteristics and generative nature of diffusion models align well with obtaining high-quality user interaction data for robust recommendation systems, while their flexible conditional generation framework can effectively integrate user intentions and preferences for accurate personalized modeling.  Based on these advantages, the ways to enhance traditional recommendations using diffusion models can be primarily categorized into three paradigms: generate high-quality interaction data, generate robust representations and preference injected conditional generation.
\par{\textbf{Generate high-quality interaction data.} Some works \cite{DGFedRS-TOIS-2025, MoDiCF-WWW-2025, Diffurec-TOIS-2025} leverage the generation capabilities of the diffusion model to generate high-quality interaction data, alleviating the data sparsity problem of the recommendation system.
DGFedRS~\cite{DGFedRS-TOIS-2025} pretrains diffusion model to capture latent personalized user information, generating high-quality interactions.
MoDiCF~\cite{MoDiCF-WWW-2025} and TDM~\cite{TDM-SIGIR-2025} focus on data missing scenario. MoDiCF~\cite{MoDiCF-WWW-2025} designs a modality-aware diffusion module to generate and iteratively optimize missing data from the learned modality-specific distribution space. TDM~\cite{TDM-SIGIR-2025} simulates extra missing data in the guidance signals and allows diffusion models to handle existing missing data through extrapolation. 
Diffurec~\cite{Diffurec-TOIS-2025} considers user/item representation as distribution and adds Gaussian noise into embedding generation in the diffusion phase to inject uncertainty and diversity.}
\par{\textbf{Generate robust representations.} \cite{ARD-SR-WWW-2025, DDRM-SIGIR-2024, DRGO-WWW-2025} utilize the denoising capability of the diffusion model to inject controlled noise into the user and item representations, and remove the noise in the reverse iterative process. 
ARD\cite{ARD-SR-WWW-2025} employs the diffusion process to refine social networks while DDRM\cite{DDRM-SIGIR-2024} and DRGO\cite{DRGO-WWW-2025} utilize diffusion model to learn robust representation.}
\par{\textbf{Preference injected conditional generation.} \cite{InDiRec-SIGIR-2025, DMCDR-KDD-2025, DimeRec-WSDM-2025} harness the information injection capabilities of diffusion models to formulate conditional probability generation tasks, thereby incorporating user intentions and preferences into the user data generation pipelie.
DMCDR\cite{DMCDR-KDD-2025} leverages the preference guidance signals from the source domain to guide the reverse process and generate the personalized user representations in the target domain.
In InDiRec\cite{InDiRec-SIGIR-2025}, conditional diffusion model is guided to generate forward views with same intent.}

\subsubsection{Target Item Generation}
The objective of recommendation aligns well with diffusion model since it infers the future interaction probabilities based on noisy historical interactions. Therefore, several works\cite{DiffRec-SIGIR-2023, DreamRec-NeurIPS-2023, DiffRIS-WWW-2024, DiQDiff-WWW-2025, DiffDiv-SIGIR-2025, DDSR-NeurIPS-2024, ADRec-KDD-2025, PreferDiff-ICLR-2025} explore diffusion models as recommenders. 

\par{\textbf{Diffusion recommender model.} DiffRec\cite{DiffRec-SIGIR-2023} treats the prediction of users' interaction as a denoising process, it further build L-DiffRec and T-DiffRec to handle large-scale item prediction and temporal modeling challenges in generative recommendtions. 
DreamRec\cite{DreamRec-NeurIPS-2023} noises the target item to explore the underlying distribution of item space, and generate recommended items directly, thereby eliminating the need for negative sampling and enabling exploration of the entire item space.
DiffRIS\cite{DiffRIS-WWW-2024}uses both local and global implicit features from user historical sequences as conditional guidance, directing the denoising process of the diffusion model to generate more personalized recommendation results.
DiQDiff\cite{DiQDiff-WWW-2025} enhances the robustness of guidance information for heterogeneous and noisy sequences through semantic vector quantization, and introduces contrastive discrepancy maximization to distinguish the denoising trajectories of different users for personalized recommendation generation.}
HorizonRec~\cite{zha2026align} proposes an align-for-fusion framework for cross-domain sequential recommendation, replacing the conventional align-then-fusion paradigm with dual-oriented diffusion models that harmonize triple-domain preferences at a fine-grained level.

\par{\textbf{Diversity and uncertainty modeling.} Although the above models have achieved competitive results, they encode user preference in the form of deterministic embeddings and rely on a homogeneous diffusion inference mechanism. This conveys a limited range of information and is insufficient to capture the diversity of user preferences.
Inspired by this, DiffDiv\cite{DiffDiv-SIGIR-2025} designs diversity-aware guided learning mechanism to guide the training of the diffusion model, enabling it to effectively capture users' diverse preferences.}

\par{\textbf{Tailored Optimization for DM-based recommendation.} DDSR\cite{DDSR-NeurIPS-2024} employs discrete diffusion to construct fuzzy sets of interaction sequences to capture the evolution of users' interests.
Some works focuse on the embedding collapse problem of applying diffusion models to recommender systems, ADRec\cite{ADRec-KDD-2025} and PreferDiff\cite{PreferDiff-ICLR-2025} argues that traditional objectives fail to leverage limits the full utilization of the generative potential of diffusion models, and propose a tailored optimization objective for diffusion based recommendaiton models.}

%% file: task.tex
\FloatBarrier
\section{Task-Level Opportunities}
\label{sec5:Task and Application-Level Opportunities}

\subsection{Top-K Recommendation}
\label{sec5.1:Next-Item Prediction}
Traditional discriminative recommendations calculate a preference score for each candidate item one-by-one, and then rank them to generate recommendations. In contrast, generative recommenders can directly generate item. However, to ensuring that the recommendation items are mapped to valid items, the generative models perform generation grounding during the inference stage. Several typical strategies have been developed:
(i) Vocabulary-Constrained Decoding restricts the decoding space of the generative model to a predefined vocabulary of item identifiers or tokens. This guarantees that every generated token corresponds to an actual item. For example, P5~\cite{P5-RecSys-2022} utilized constrained decoding with a predefined item vocabulary and applied beam-search, ensuring outputs correspond to valid catalog items. 
IDGenRec~\cite{IDGenRec-SIGIR-2024} utilized a prefix tree to store all generated candidate
IDs. Each newly generated token is constrained by the previously
generated tokens, ensuring that the generation process only considers tokens that can potentially form an existing candidate ID
in the dataset.
\cite{RecSysLLM-arxiv-2023} and \cite{IndexItem-SIGIRAP-2023} proposed
to utilize the Trie algorithem for constrained generation, where the generated identifier is guaranteed to be a valid identifier. However, Trie strictly generates the valid identifier from the first token, where the accuracy of the recommendation depends highly on the accuracy of the
first several generated tokens. To combat this issue, TransRec~\cite{TransRec-KDD-2024}
utilized FM-index to achieve position-free constrained generation,
which allows the generated token from any position of the valid
identifier. The generated valid tokens will then be grounded to the
valid identifiers through aggregations from different views. TransRec also introduced multi-facet identifiers (ID, title, attributes) and an Aggregated Grounding Module to map the generated identifiers back to in-corpus items.
(ii) Post-Generation Filtering. Post-generation filtering lets an LLM freely generate text (IDs, titles, or semantic tokens) and then maps/reranks those outputs to in-catalog items via exact/semantic matching or reranking. For example, BIGRec~\cite{BIGRec-TORS-2025} proposed to
ground the generated identifier to the valid items via L2 distance
between the generated token sequences’ representations and the
item representation. 
These approaches are efficient and scalable, but heavily depend on the quality of item embeddings.
(iii) Prompt Augmentation. For LLM-based recommenders, this strategy injects the candidate item set into text prompts and asks the model to recommend items from the candidate set. Many LLM-based recommenders adopt this strategy, such as LLaRA~\cite{LLaRA-SIGIR-2024}, A-LLMRec~\cite{A-LLMRec-KDD-2024}, iLoRA~\cite{iLoRA-NeurIPS-2024}.

\subsection{Personalized Content Generation}
\label{sec5.2:Recommended Item Generation}
In addition to the typical recommendation that requires valid
generation to recommend existing items to users, another
research direction capitalizes on the generative capabilities of models to create entirely new content tailored to individual users. It is important to clarify how this direction differs from general AI-generated content (AIGC): while general AIGC focuses on open-ended content creation (e.g., text-to-image synthesis from arbitrary prompts), personalized content generation in recommendation is fundamentally preference-conditioned and task-situated. Specifically, the generation process is guided by the personalized user preferences, and its primary objective is to enhance the recommendation experience, rather than producing standalone creative works. We organize existing efforts into personalized visual content generation and personalized textual content generation. (i) Personalized visual content generation. Many Recent works leverage diffusion models to generate personalized recommendation content.
For example, DiFashion~\cite{DiFashion-SIGIR-2024} generates personalized outfit combinations based on a user's style preferences, which can serve both as direct recommendations and as guidance for fashion production.
With the advancement of large-scale model technologies, recent research work focuses on leveraging diffusion models to generate virtual visualizations of recommendation results within the recommendation process. DreamVTON~\cite{Dreamvton-MM-2024} addresses the multi-view consistency problem in personalized diffusion models for 3D generation through a template-driven optimization mechanism and normal-style LoRA, achieving high-quality 3D virtual try-on based solely on person images, garment images, and text prompts. InstantBooth~\cite{Instantbooth-CVPR-2024} introduces a concept encoder and patch encoder to learn global concept embeddings and rich local patch features respectively, and incorporates novel adapter layers and concept token normalization techniques to enable personalized image generation. OOTDiffusion~\cite{Ootdiffusion-AAAI-2025} designs an outfitting UNet to learn detailed garment features and employs outfitting fusion to precisely align garment features with the human body in self-attention layers, while introducing outfitting dropout to implement classifier-free guidance, thereby generating high-quality, controllable virtual try-on images without requiring explicit deformation processes.
(ii) Personalized textual content generation~\cite{PersonalizedGenSurvey-ACL-2025}. Some works leveraged realistic user interaction to explore personalization for review generation~\cite{Justifying-EMNLPIJCNLP-2019,TemplateExplanation-CIKM-2020,DualLearning-WWW-2020,PETER-ACL-2021} and news headline generation~\cite{PENS-ACL-2021,Generating-ACL-2023,General-TACL-2023}. For
example, Ao et al. \cite{PENS-ACL-2021} presents a personalized
headline generation benchmark by collecting user’s
click history on Microsoft News.

As generative models become increasingly capable, the boundary between selecting existing items and creating new personalized content is progressively blurring. This convergence positions personalized content generation as a distinctive frontier of generative recommendation.

\subsection{Conversational Recommendation}
\label{sec5.3:Conversational Recommendation}
Conversational recommendation is capable of eliciting the dynamic preferences of users and taking actions based on their current needs through real-time multi-turn interactions using natural language.
The previous works mostly focus on single-turn LLM recommendation personalized based on pre-existing user history and item text. Nowadays, LLMs provide new opportunities for multi-turn conversational recommender systems (CRSs) where each turn presents a chance for the user to clarify or revise their preferences, critique and ask questions about recommended items, or convey a variety of other real-time intents. 

In this subsection, we provide a taxonomy of recent research on LLM-based conversational recommendation, categorizing them into five methodological directions.
(i) Prompting and Zero-shot Methods. One of the earliest applications of LLMs to CRS is through prompting, where task-specific templates or demonstration examples are crafted to steer LLMs toward generating recommendations. These methods often exploit the zero-shot or few-shot capabilities of LLMs. For instance, He et al., \cite{zeroshotCRS-CIKM-2023} showed that off-the-shelf LLMs can outperform supervised CRS baselines without fine-tuning. Later studies such as \cite{chatgptCRS-Arxiv-2024} incorporated user feedback iteratively into prompts, improving personalization over multiple turns. \cite{DCRS-SIGIR-2024} highlighted the effectiveness of combining demonstrations or structured contexts to guide the model. \cite{CP-Rec-AAAI-2023} structured dialogue context into prompts for better recommendations.
(ii) Retrieval-augmented and Knowledge-enhanced Approaches. A major limitation of purely prompt-based methods is the lack of grounding in the item space, which can lead to hallucinations. To address this, recent works combine LLMs with retrieval modules or knowledge graphs. For example, \cite{GraphRAGCRS-PAKDD-2025} enhanced recommendations by retrieving relevant items and entities. \cite{RetrievalCRS-RecSys-2024} integrated collaborative filtering signals into the retrieval process, while \cite{KGPLM-CRS-TNNLS-2025} focused on jointly modeling semantic and structural information. These retrieval-enhanced approaches improve factual grounding and personalization but come at the cost of higher latency and knowledge maintenance overhead.
(iii) Beyond prompting and retrieval, researchers have developed unified and parameter-efficient architectures. \cite{ravaut-etal-2024-parameter} reformulated CRS as a single NLP task. MemoCRS~\cite{MemoCRS-CIKM-2024} introduced memory modules to capture sequential coherence, while \cite{ILMCRS-Arxiv-2025} represented items directly in natural language. In parallel, Chat-REC~\cite{Chat-REC-2023} enhanced interaction and explainability.
(iv) Evaluation has also received attention. \cite{BehaviorAlignment-SIGIR-2024} proposed assessing whether system strategies align with human expectations rather than relying only on accuracy metrics. This line of work points to the need for new evaluation protocols, bias mitigation, and human-in-the-loop assessments.

\subsection{Explainable Recommendation}
\label{sec5.4:Explainable Recommendation}
Explainable recommendation has demonstrated
significant advantages in informing users about the logic behind recommendations, thereby increasing system transparency, effectiveness, and trustworthiness. Early works focus on generating explanations with predefined templates or extracting logic reasoning rules from recommendation models. With the advance of LLMs, a burgeoning body of research started to build LLM-based explainable recommendation models.  (i) P5~\cite{P5-RecSys-2022} mainly focus on designing prompts to guide LLMs to directly generate explanations. However, LLMs face the hallucination problem, resulting in generating low-quality explanations. To address these challenges, LLM2ER~\cite{LLM2ER-AAAI-2024} fine-tuned LLM-based explainable recommendation backbone in a reinforcement learning paradigm with two explainable quality reward models.
(ii) Recently, researchers have tried to combine the advantages of graphs to enhance the explanations generated by LLMs.
XRec~\cite{XRec-EMNLP-2024} adopts a graph neural network (GNN) to model the graph structure and generate embeddings. Then, the embeddings are fed into LLMs to generate explanations. G-Refer~\cite{G-Rec-WWW-2025} leveraged a hybrid graph retrieval to extract explicit CF signals, employing knowledge pruning to filter out less relevant samples, and utilizing retrieval-augmented fine-tuning to integrate retrieved knowledge into explanation generation. 
(iii) Some existing works~\cite{ThinkRec-Arxiv-2025,Reason-to-Recommend-Arxiv-2025} on large model-based recommendation leverage the reasoning abilities of thinking models and use the thought process as an explanation for recommendations. The challenge in this type of work lies in how to construct the ground truth for the explanation.

\subsection{Recommendation Reasoning}
\label{sec5.5:Recommendation Reasoning}
Recent advances in LLMs have led to remarkable progress in reasoning capabilities, with representative models including DeepSeek-R1~\cite{DeepSeekR1-Nature-2025}, GPT o-series, etc. LLMs achieve complex reasoning through various prompting techniques, with CoT prompting~\cite{CoT-NIPS-2022} serving as the foundational approach that decomposes problems into intermediate reasoning steps. This has inspired numerous extensions, including zero-shot CoT~\cite{ZeroShotCoT-NIPS-2022}, self-consistency decoding~\cite{SelfConsistency-ICLR-2023}, and tree-of-thoughts~\cite{TreeOfThought-NIPS-2023}. These techniques enable test-time scaling where additional computational budget during inference improves performance. Recent work has shifted focus from prompting to post-training enhancement of reasoning capabilities by using Reinforcement Learning techniques.
Reasoning capability has recently garnered significant attention in RSs research. Reasoning-based RSs aim to perform multi-step deduction to deliver more accurate and explainable recommendations.
Existing studies can be broadly grouped into three categories.
(i) Explicit reasoning methods generate explicit and human-readable reasoning processes. Reason4Rec~\cite{Reason4Rec-Arxiv-2025} introduced the task of deliberative recommendation, which requires LLMs to explicitly reason before predicting user feedback. It further develops a multi-expert framework consisting of preference distillation, preference matching, and feedback prediction. Reason-to-Recommend~\cite{Reason-to-Recommend-Arxiv-2025} proposed Interaction-of-Thought (IoT) reasoning, which organized user–item interaction chains into structured reasoning paths and trains LLMs via a two-stage SFT and RL paradigm. Similarly, ThinkRec~\cite{ThinkRec-Arxiv-2025} leveraged specially designed thinking prompts to induce LLMs to generate reasoning chains that enhance both recommendation quality and explainability. These works highlight the benefits of making the reasoning process explicit and transparent.
OneRec-Think~\cite{OneRec-Think-Arxiv-2025} activated the LLM-based recommender's reasoning capabilities through CoT-based SFT. To generate effective CoT examples, the framework first extracted coherent reasoning trajectories from pruned user contexts, then leveraged these trajectories to guide CoT generation over raw behavioral data, enabling effective contextual distillation for noisy industrial settings. Finally, OneRec-Think optimized reasoning quality using recommendation-specific RL reward functions.
(ii) Implicit reasoning methods perform latent reasoning without textual interpretability. LatentR³~\cite{LatantR3-Arxiv-2025} introduced reinforced latent reasoning, which encodes reasoning processes into a compact sequence of latent tokens rather than generating lengthy textual CoTs. Through supervised fine-tuning followed by RL optimization, LatentR³ achieves both efficiency and effectiveness, offering a practical paradigm for large-scale online recommendation scenarios. ReaRec~\cite{ReaRec-Arxiv-2025} introduced an inference-time computing framework that enabled traditional sequential recommenders to perform multi-step latent reasoning by autoregressively processing hidden states with reasoning position embedding. STREAM-Rec~\cite{Stream-Rec-Arxiv-2025} introduced a slow thinking paradigm for sequential recommendation, enabling traditional lightweight recommenders to perform multi-step deliberative reasoning rather than one-step direct matching. The core innovation is an iterative residual-based reasoning method, where the model progressively refines predictions by computing and fitting the residual between its current output and the target behavior representation, with these intermediate residual corrections serving as reasoning paths. 
(iii) LLM reasoning augmentation methods leveraged LLMs to generate reasoning steps that enhance the training of traditional RSs. For example, DeepRec~\cite{DeepRec-Arxiv-2025} proposed an autonomous interaction paradigm where the LLM hypothesizes user preferences, queries a traditional RS for candidates, and iteratively refines its reasoning before final recommendation. LLMRG~\cite{LLMRG-AAAI-2024} constructed reasoning graphs through LLM-driven chain reasoning, divergent extension, and self-verification, and then integrated them into recommendation models via graph neural networks.

%% file: challenges.tex
\section{Discussion and Open Challenges}
\label{sec6:Open Challenges}
\begin{figure*}[!t]
    \centering \includegraphics[width=0.98\textwidth]{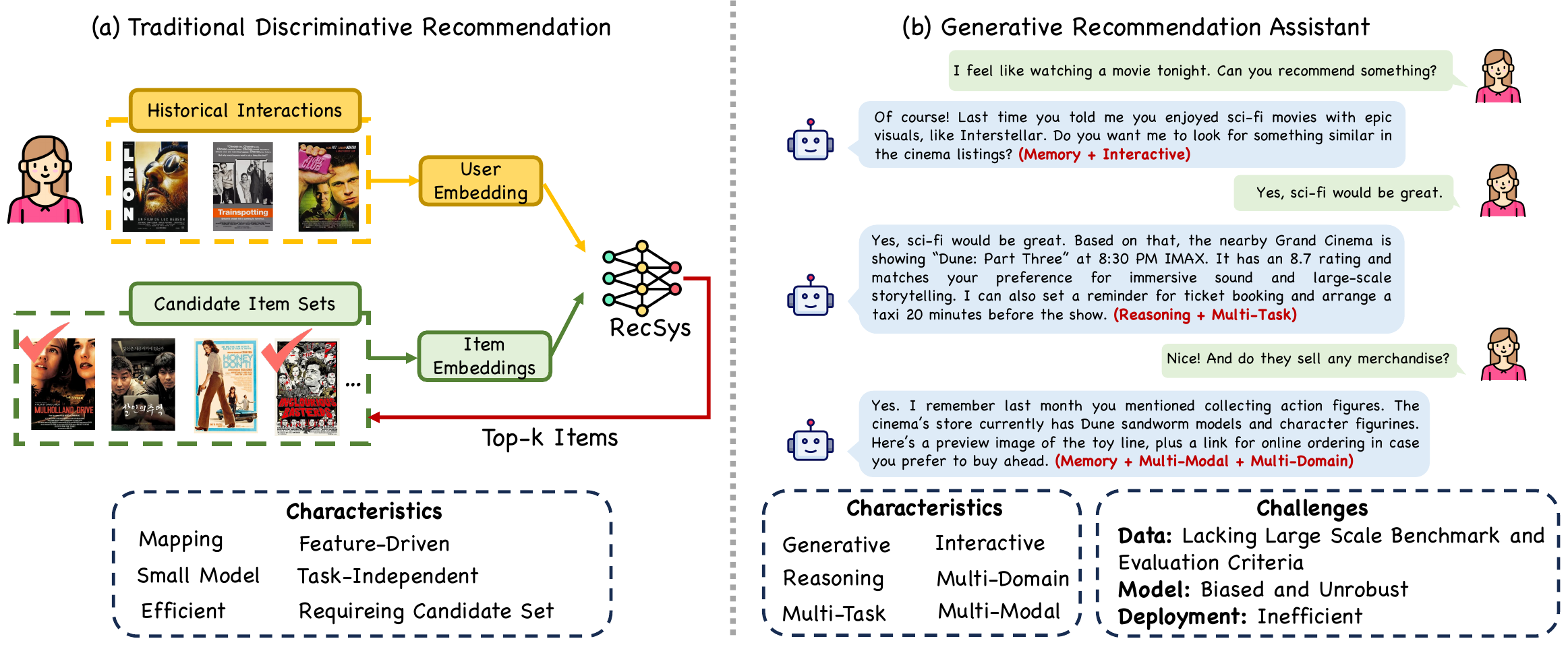}
    \caption{Illustration of traditional discriminative recommendation and generative recommendation assistant.}
    \label{fig:challenges}
\end{figure*}
\subsection{Discussion}
In the previous sections, we summarized current generative recommendation methods from the aspects of data, model, and task. We can observe that the emergence of generative models has led to a paradigm shift in RSs. This shift makes the field more open and flexible across all three dimensions: data, model, and task.

Specifically, data-wise, traditional discriminative RSs relied on hand-crafted features from domain-specific datasets. These features were manually curated for one specific and fixed task. Generative models, particularly LLMs, bring world knowledge into the recommendation process. 
Additionally, LLMs' ability to perform real-time internet searches introduces new opportunities for bringing massive, up-to-date world knowledge. Moreover, LLMs are naturally suited for integrating recommendation knowledge from multi-domain and tasks. In summary, generative models bring more knowledge to the data side.

On the model side, generative models have demonstrated the power of scaling laws. As model size, training data, and computational resources increase, recommendation performance improves significantly. Larger models also exhibit emergent capabilities. For example, instruction fine-tuning, RLHF, and in-context learning changes the way of training and inference. Reasoning ability enables the model to understand and infer complex relationships between users, items, and contexts in ways that smaller traditional models cannot. Moreover, unlike traditional discriminative recommenders that restrict a candidate set, generative models are capable of generating recommendations directly. In summary, large generative models bring stronger capabilities to the model side.

On the task side, traditional RSs were often designed to perform a single task, such as CTR prediction or rating prediction. While LLMs are task-agnostic by design, meaning they can handle a variety of recommendation tasks within a single framework. The same model can rank items, generate explanations, and produce personalized content in an interactive way. This flexibility allows the system to adapt dynamically to different user needs. In short, generative recommendation supports more diverse and flexible functionalities at the task dimension.

Generative RSs have broken away from the traditional discriminative paradigm, which were mapping-based, feature-driven, small-model, task-independent, and reliant on pre-defined candidate sets. Nowadays, generative recommenders are reshaping the recommendation paradigm. They are moving towards recommendation assistants that are open, capable of handling a wide range of tasks, and able to adapt to changing user needs. These systems can interact with users more flexibly, supporting multiple tasks such as ranking, explanation, content generation, and conversation. This shift represents a fundamental change in how recommendations are made, making the process more dynamic and responsive to diverse user requirements. The difference between the two paradigms is shown in Figure \ref{fig:challenges}. 

While the paradigm shift from traditional recommendation systems to generative models opens up new possibilities, there is still a long way to go before realizing the full potential of recommendation assistants. The journey from the current state to the future vision requires overcoming several challenges. In the following sections, we will explore these challenges across three key dimensions: data, model, and deployment.

\begin{table*}[ht]
\centering
\footnotesize
\caption{Representative evaluation metrics for generative recommender systems, grouped by evaluation scenario.}
\label{tab:metrics}
\renewcommand{\arraystretch}{1.8}
\setlength{\tabcolsep}{8pt}
\begin{tabular}{|>{\centering\arraybackslash}m{1.2cm}|>{\centering\arraybackslash}m{1.45cm}|>{\centering\arraybackslash}m{4.85cm}|m{7cm}|}
\hline
\textbf{Category} & \textbf{Metric} & \textbf{Formulation} & \multicolumn{1}{>{\centering\arraybackslash}m{7cm}|}{\textbf{Notes}} \\
\hline

\multirow[c]{8}{*}{\parbox{1.3cm}{\centering Recom-\\mendation}}
  & NDCG@K
  & $\dfrac{\sum_{i=1}^{K} rel_i / \log_2(i+1)}{\sum_{i=1}^{K} rel_i^* / \log_2(i+1)}$
  & Position-aware ranking metric; $rel_i \in \{0,1\}$, $rel_i^*$ is the ideal ranking. Penalizes relevant items ranked lower. \\
\cline{2-4}
  & Recall@K
  & $\dfrac{\sum_{i=1}^{K} rel_i}{|\mathcal{G}|}$
  & Fraction of ground-truth items $\mathcal{G}$ retrieved within top-$K$. \\
\cline{2-4}
  & Precision@K
  & $\dfrac{\sum_{i=1}^{K} rel_i}{K}$
  & Fraction of top-$K$ items that are relevant. Complements Recall@K by penalizing irrelevant recommendations. \\
\cline{2-4}
  & MRR@K
  & $\dfrac{1}{|\mathcal{U}|}\displaystyle\sum_{u\in\mathcal{U}}\dfrac{1}{\mathrm{rank}_u^*}$
  & Mean Reciprocal Rank; $\mathrm{rank}_u^*$ is the rank of the first relevant item for user $u$ within top-$K$. \\
\cline{2-4}
  & HR@K
  & $\mathbf{1}\!\left(\sum_{i=1}^{K} rel_i > 0\right)$
  & Binary indicator of whether at least one relevant item appears in top-$K$; $\mathbf{1}(\cdot)$ is the indicator function. \\
\cline{2-4}
  & AUC
  & $\dfrac{\sum_{i\in\mathcal{P}}\sum_{j\in\mathcal{N}}\mathbf{1}[s_i > s_j]}{|\mathcal{P}||\mathcal{N}|}$
  & $\mathcal{P}$/$\mathcal{N}$: positive/negative item sets; $s_i,s_j$: predicted scores. Threshold-independent and robust to class imbalance. \\
\cline{2-4}
  & CTR
  & $\dfrac{\text{\# Clicks}}{\text{\# Impressions}}$
  & Click-through rate; most direct signal of short-term user engagement. \\
\cline{2-4}
  & CVR
  & $\dfrac{\text{\# Conversions}}{\text{\# Clicks}}$
  & Conversion rate from click to target action (e.g., purchase); captures downstream business value. \\
\hline

\multirow[c]{5}{*}{\parbox{1.3cm}{\centering Content Gen. Quality}}
  & BLEU
  & $\begin{cases} \frac{|S\cap R|}{|S|}\cdot\exp\!\left(1-\frac{r}{c}\right) & c < r \\[2pt] \frac{|S\cap R|}{|S|} & c \geq r \end{cases}$
  & $S$: generated text, $R$: reference, $c$/$r$: their lengths. Includes a brevity penalty when $c < r$. \\
\cline{2-4}
  & ROUGE-L
  & $\dfrac{LCS(S,R)}{|R|}$
  & Longest common subsequence between $S$ and $R$ normalized by $|R|$. Captures structural similarity beyond n-gram matching. \\
\cline{2-4}
  & SBERT
  & $\dfrac{\mathbf{v}_S\cdot\mathbf{v}_R}{\|\mathbf{v}_S\|\cdot\|\mathbf{v}_R\|}$
  & Cosine similarity of sentence embeddings $\mathbf{v}_S$, $\mathbf{v}_R$. Captures semantic similarity without requiring lexical overlap. \\
\cline{2-4}
  & LLM-E
  & $\text{LLM}(S, R)$
  & An LLM judge scores $S$ against $R$ on relevance, fluency, and factuality. \\
\cline{2-4}
  & FID
  & $\|\mu_r{-}\mu_g\|^2 + \mathrm{Tr}\!\left(\Sigma_r{+}\Sigma_g{-}2(\Sigma_r\Sigma_g)^{1/2}\right)$
  & Fr\'{e}chet Inception Distance for visual content; $(\mu_r,\Sigma_r)$/$(\mu_g,\Sigma_g)$: Inception feature statistics of real/generated images. Lower is better. \\
\hline

\multirow[c]{3}{*}{Diversity}
  & ILD
  & $\dfrac{1}{K(K-1)}\displaystyle\sum_{\substack{i,j=1\\i\neq j}}^{K} d(\mathbf{v}_i,\mathbf{v}_j)$
  & Intra-List Diversity; $d(\cdot,\cdot)$ is embedding distance between items in the list. Higher values indicate more diverse recommendations. \\
\cline{2-4}
  & Coverage
  & $\dfrac{\left|\bigcup_{u\in\mathcal{U}}\mathcal{L}_u\right|}{|\mathcal{I}|}$
  & Fraction of catalog $\mathcal{I}$ covered by recommendation lists $\{\mathcal{L}_u\}$. Measures ability to surface long-tail items. \\
\cline{2-4}
  & Novelty
  & $\dfrac{1}{K}\displaystyle\sum_{i=1}^{K} -\log_2 p(i)$
  & $p(i)$: popularity of item $i$ in training data. Rewards for recommending less-known items. \\
\hline

\multirow[c]{2}{*}{Fairness}
  & DP
  & $\max_{g,g'\in\mathcal{G}}\!\left|\,P_g(\hat{y}{=}1) - P_{g'}(\hat{y}{=}1)\,\right|$
  & Demographic Parity gap across sensitive groups $\mathcal{G}$ (e.g., gender, age). Zero indicates equal recommendation rates across groups. \\
\cline{2-4}
  & EO
  & $\max_{g,g'\in\mathcal{G}}\!\left|\,P_g(\hat{y}{=}1\mid y{=}1) - P_{g'}(\hat{y}{=}1\mid y{=}1)\,\right|$
  & Equal Opportunity gap; compares true positive rates across sensitive groups. Zero indicates relevant items are surfaced equally regardless of group. \\
\hline

\multirow[c]{2}{*}{\parbox{1.3cm}{\centering Conv. Rec.}}
  & SR
  & $\dfrac{1}{|\mathcal{D}|}\displaystyle\sum_{d\in\mathcal{D}}\mathbf{1}(\text{success}_d)$
  & Success Rate over dialogues $\mathcal{D}$; $\text{success}_d$ indicates a relevant item was recommended within the allotted turns. \\
\cline{2-4}
  & AT
  & $\dfrac{1}{|\mathcal{D}|}\displaystyle\sum_{d\in\mathcal{D}} T_d$
  & Average Turns to success; $T_d$ is the turn count of session $d$. Fewer turns indicate more efficient preference elicitation. \\
\hline

\end{tabular}
\end{table*}

\subsection{Open Challenges}
\subsubsection{Data}
\textbf{Dataset.}
As a data-driven and user-centric research direction, the development of RS has always been closely tied to the availability and evolution of datasets. Key datasets have driven significant advancements in the field, providing benchmarks for training and evaluating algorithms and spurring new research directions.
The MovieLens dataset played a foundational role in the early stage, providing large-scale rating data that enabled the development and evaluation of collaborative filtering and matrix factorization methods. Later, the release of the Netflix Prize dataset became a milestone that pushed forward latent factor models and matrix factorization techniques, demonstrating their effectiveness in large-scale recommendation tasks. The Amazon Review dataset further shifted attention from explicit ratings to implicit feedback such as clicks and purchases, and its combination of ratings, text, and product metadata fostered research on hybrid recommendation that integrates multiple signals. The Yelp dataset, with its rich user reviews and business information, advanced the use of deep learning-based recommendation, where natural language processing was leveraged for sentiment analysis, representation learning, and context-aware suggestions.

These datasets have greatly advanced the development
of recommendation systems, serving as the foundation for
many breakthroughs over the past two decades. However,
for generative recommendation, they are no longer fully
suitable.Most of them are non-interactive,offline, and static,
 capturing only point-in-time user preferences rather than the
dynamic feedback loops and multi-round interactions that
characterize real-world scenarios. On top of these datasets,
most existing studies rely on fixed task-specific prompt
templates to generate recommendations and use traditional
metrics for evaluation. These datasets are more suitable
for assessing the accuracy performance of traditional RSs,
but they restrict the assessment of generative models as
personalized assistants, which in reality need to operate across multiple scenarios and handle diverse tasks in interactive settings. This gap highlights the urgent need for
new benchmarks that can better support the next stage of
recommendation research.

\textbf{Evaluation Metrics.} 
To provide a clearer picture of the current evaluation landscape, Table \ref{tab:metrics} summarizes representative metrics used across different generative recommendation scenarios. These metrics can be broadly grouped into five categories: ranking-oriented metrics (e.g., NDCG@K, Recall@K, HR@K) that assess the accuracy of top-K item retrieval; content generation quality metrics (e.g., BLEU, ROUGE-L, SBERT, LLM-E, FID) that evaluate the fluency, semantic fidelity, and visual realism of generated content; diversity metrics (e.g., ILD, Coverage, Novelty) that measure the breadth and freshness of recommendation lists; fairness metrics (e.g., DP, EO) that quantify demographic parity across sensitive user groups; and conversational metrics (e.g., Success Rate, Average Turns) that capture the efficiency of preference elicitation in multi-turn dialogues.
While these metrics collectively cover a wide range of evaluation dimensions, they each carry inherent limitations when applied to generative recommendation systems. Ranking metrics such as NDCG@K and HR@K presuppose a fixed candidate set and binary relevance labels, making them ill-suited for open-ended generation scenarios where the output is not constrained to a predefined item pool. Content generation metrics like BLEU and ROUGE-L rely on lexical overlap with reference texts, which poorly reflects semantic quality in free-form personalized generation. Although LLM-based evaluation (LLM-E) offers a more flexible alternative, it introduces non-determinism and incurs substantial computational cost at scale. Diversity and fairness metrics, while important, are rarely jointly optimized or reported alongside accuracy metrics in existing benchmarks, leaving a gap in holistic evaluation. Conversational metrics such as Success Rate and Average Turns depend heavily on the design of dialogue simulators, whose fidelity to real user behavior remains questionable. 

Beyond these metrics, the agent-based simulation works reviewed in Section~\ref{sec3.1.2:Agent-Based Behavior Simulation} open up new possibilities for recommendation evaluation: by generating diverse and controllable interaction signals, they offer a promising alternative to static offline benchmarks, enabling dynamic, multi-round, and cost-effective evaluation that better reflects real-world recommendation scenarios. However, despite this promise, their validity as behavioral proxies remains an open question. On one hand, the behavioral fidelity of LLM-based agents is inherently constrained by a chicken-and-egg dilemma: building accurate user simulators requires large-scale real user behavior data, yet the very motivation for simulation is to compensate for the lack of such data. As a result, simulated behaviors risk reflecting the distributional biases of pretraining corpora rather than the true diversity of real user preferences. On the other hand, most existing simulation frameworks model users as static agents with fixed personas, failing to capture the dynamic nature of real user behavior, such as preference drift, mood-driven decisions, and evolving interests over time.
Taken together, the fragmented and task-specific nature of existing evaluation protocols makes it difficult to assess generative recommendation systems as unified, multi-capable assistants. This underscores the urgent need for new benchmarks that support dynamic, multi-task, and interactive evaluation aligned with the full potential of generative models.

\subsubsection{Model}
At the model level, generative RSs face significant challenges in terms of bias and robustness.

\textbf{Bias.} We categorize the main biases in existing generative RSs into three types.
(i) The first is popularity bias. The training data bias in generative recommendation primarily arises from two aspects: the biases present in large-scale pretraining corpora and those found in sparse user interaction. The ranking results of LLMs are influenced by the popularity of items in user interaction data, and popular items that are frequently discussed and mentioned in the LLM’s pretraining corpus tend to be ranked higher. This may lead to a lack of diversity in the responses and potentially marginalize less popular or minority viewpoints.
Existing methods~\cite{FAIRREC-SIGIR-2025, Flower-SIGIR-2025, sprec-WWW-2025,IFairLRS-WWW-2024, SERAL-KDD-2025}~mitigate the fairness issue from the perspective of adjusting or generating training data distribution. Although effective, further research is needed in utilizing the generative capabilities of large models to design prompts that extract or generate fair user interactions.
Additionally, existing training methods may exacerbate popularity bias~\cite{bifair-ARXIV-2025}. SFT maximizes the likelihood function, is prone to overfitting on popular items, thereby amplifying popularity bias and reducing the diversity of recommendation results. DPO, on the other hand, tends to suppress the probability of non-preferred responses, reinforcing existing patterns in the data while being highly sensitive to the quality of preference pairs. (ii) The second is fairness. Generative model exhibits fairness issues related to sensitive attributes~\cite{invdiff}, as the model implicitly utilizes demographic characteristics of individuals involved in certain task annotations from user interaction data or pretraining data. This may lead the model to make recommendations based on assumptions that users belong to specific groups~\cite{up5-EACL-2024, xu2025fair, LLMFOSA-ARXIV-2025}, such as biases in recommendations due to gender or race. Addressing these fairness issues is critical and necessary to ensure fair and unbiased recommendations. Determining whether the model can utilize or infer sensitive attributes of users, and reducing the overlap between user preference modeling and the representation of sensitive user information, is key to mitigating fairness issues on the user side.
(iii) The third is position bias. 
In the generative recommendation modeling paradigm, user behavior sequences and recommended candidate items are input into the language model in the form of text sequences, which may introduce certain positional biases inherent to the language model itself~\cite{kweon2025uncertainty-WWW-2025, ganprompt-ARXIV-2024}. LLMs are highly sensitive to the structure and content of input prompts, with even minor changes potentially leading to significant differences in the output. Variations in prompt design, such as the complexity of product descriptions, the number of candidate items, or their order, can influence the model's attention distribution, thereby increasing the uncertainty of the prompt. For instance, LLMs typically prioritize higher-ranked items.

\textbf{Robustness.} For robustness, we categorize the main challenges into robustness against natural noise and robustness against malicious attack. (i) For robustness against natural noise, recommendation tasks have long been plagued by noise issues, such as user interactions being influenced by clickbait or disrupted by unintended interactions. The challenge of denoising lies in the lack of clear noise signals, with existing denoising methods largely relying on joint training guided by the recommendation objective~\cite{IDEA-SIGIR-2025, SGIL-SIGIR-2025}. With the rise of generative recommendation, we aim to leverage the built-in robustness of LLMs, including their extensive open knowledge and advanced semantic reasoning capabilities, to identify noisy interactions and provide more reliable suggestions~\cite{LLM4DSR-TOIS-2024, LLaRD-WWW-2025, DALR-TOIS-2025, LLMHD-ARXIV-2024, RuleAgent-arxiv-2025}. However, directly applying LLMs to denoising presents several challenges: there is a significant gap between the pretraining objectives of LLMs and the specific requirements of recommendation denoising, and direct denoising prompts often result in completely incorrect noise classification outcomes. Even with fine-tuning, LLMs can summarize user preferences from interaction items, but the presence of false positive samples in historical interactions can cause biases in the summary of user preferences, thus reducing the effectiveness of noise sample identification~\cite{LLMHD-ARXIV-2024, RuleAgent-arxiv-2025}. Additionally, LLMs are prone to the notorious hallucination phenomenon, where they may incorrectly label non-existent interactions as noise. Therefore, it is essential to consider how to constrain the knowledge generated by LLMs to align with the specific prediction objectives in recommendation tasks.
(ii) For robustness against malicious attack~\cite{yang2026revisiting,liu2026wdpo}, injection attacks represent a competitive approach to attacking traditional recommender systems~\cite{nguyen2024manipulating, gunes2014shilling}.
While effective, it is costly due to the need to inject a substantial number of malicious profiles. Such large-scale injections are infeasible due to limited attack budgets~\cite{id-sigir-2025}. In contrast, due to the sensitivity of LLMs to text description, textual simulation attack is a potential direction. This attack directly rephrases the original item description of target items into an untrue crafted version. The adversarial texts are semantically similar to the originals, and align with the stylistic patterns of the system, all while exerting minimal impact on overall recommendation performance~\cite{id-sigir-2025, zhang2024stealthy-ACL-2024, CheatAgent-KDD-2024, retrieval-ARXIV-2025}. Compared to traditional recommendation attacks, textual attacks are remarkably low-cost, and can be executed under black-box settings, lowering the barrier to attack and raising the practical risk. Moreover, the malicious texts generated through such attacks exhibit transferability across different LLM-based recommendation models and tasks, meaning that adversarial prompts tailored to a single model may effectively compromise multiple systems or applications.
As researchers increasingly focus on multi-modal LLM-based generative recommendations~\cite{harnessing-AAAI-2025, notellm-KDD-2025}, the robustness of these models against attacks has emerged as a potential research direction. Furthermore, although studies on attacks targeting LLM-based generative recommendations are still in their preliminary stages, existing defense strategies exhibit limited effectiveness against data poisoning attacks~\cite{id-sigir-2025}. 

\subsubsection{Deployment}
Beyond the challenges at the data and the model level, deploying generative RSs in real-world scenarios introduces difficulties of efficiency. Unlike traditional recommender models, generative models are usually large and computationally intensive, which makes their deployment far from trivial. 

\textbf{Training Efficiency.}
Although parameter-efficient fine-tuning (PEFT) methods reduce the training cost of LLM-based generative recommendations by updating only a subset of model parameters, they remain insufficient to address the challenges posed by the rapidly increasing scale of recommendation datasets. The key to improving fine-tuning efficiency lies in rapidly adapting large language models to perform effectively in recommendation tasks using fewer data and computational resources. One intuitive approach is to select a small, representative subset of data ~\cite{DEALRec-sigir-2024,GORACS-kdd-2025}. However, such methods typically require computing the influence or gradient information for each sample, which incurs significant computational complexity. Moreover, selections based on individual interactions often overlook the correlations between interaction records in recommendation tasks, leading to suboptimal results.

\textbf{Inference Efficiency.}
Generative recommendation faces a significant issue of inefficient inference: its inference process, due to the nature of autoregressive decoding, requires multiple serial calls to the LLM, resulting in excessive time consumption, which hinders its practical application in real-time recommendation scenarios. How can we leverage the superior performance of LLM-based recommenders while maintaining inference latency as low as traditional recommenders? Unlike natural language processing (NLP) tasks, recommendation tasks require the generation of the top-K distinct sequences (recommended item lists), which necessitates beam search during the decoding process to ensure the generation of diverse and high-quality recommendation results. This makes traditional language model inference acceleration methods, such as speculative decoding, difficult to apply directly~\cite{AtSpeed-ICLR-2025}. Knowledge distillation~\cite{DLLM2Rec-recsys-2024, SLMREC-ICLR-2025} alleviates this issue to some extent by transferring knowledge from a complex LLM-based recommendation model to a lightweight traditional recommender or a smaller language model.

%% file: conclusion.tex
\section{Conclusion}
\label{sec7:Conclusion}
This survey has systematically examined how generative models are revolutionizing recommender systems, marking a paradigm shift from discriminative matching to intelligent synthesis.
Our investigation reveals a tripartite transformation. At the data level, generative models transcend traditional boundaries through knowledge augmentation and behavioral simulation. At the model level, LLM-based methods, large-scale architectures, and diffusion approaches establish powerful alternatives to conventional paradigms. At the task level, new capabilities emerge including conversational dynamics, transparent reasoning, and personalized content creation, fundamentally redefining user-system interaction.
Critical challenges remain: developing dynamic benchmarks that capture real-world complexity, fortifying models against biases and attacks, and achieving deployment efficiency.
We envision recommendation systems evolving into cognitive assistants that are transparent, contextually aware, and seamlessly integrate reasoning and generation through natural language. This represents not incremental improvement but a fundamental reconceptualization of how intelligent systems serve human information needs.
As we stand at this inflection point, this survey serves as both a comprehensive synthesis of current achievements and a strategic roadmap for future research, guiding the community toward realizing the full potential of generative intelligence in recommendation systems.

%% file: acknowledgment.tex
\section{Acknowledgments}
This work was supported in part by grants from the National Natural Science Foundation of China (Grant No. U23B2031, 62436003, and 62402159), the Fundamental Research Funds for the Central Universities (Grant No. JZ2025HGPB0248).